\documentclass[10pt,journal,compsoc]{IEEEtran}
\usepackage{graphicx}
\usepackage{amsmath}
\usepackage{amssymb}
\usepackage{amsthm}
\usepackage{afterpage}
\usepackage{verbatim}
\usepackage{psfrag}
\usepackage{color}
\usepackage{cite}
\usepackage{setspace}
\usepackage{epsfig}
\usepackage{epstopdf}
\usepackage{footmisc}
\usepackage{fmtcount}
\usepackage{stackrel}
\usepackage{float}
\usepackage{breqn}
\usepackage{enumerate}
\usepackage{graphicx}
\usepackage{subcaption}
\usepackage{algpseudocode,algorithm,algorithmicx}
\usepackage{graphicx}
\usepackage{ragged2e}

\begin{document}
\title{Performance Analysis of Connectivity and Localization in Multi-Hop Underwater Optical Wireless Sensor Networks}
\author{ Nasir Saeed,~\IEEEmembership{Member,~IEEE}, Abdulkadir Celik,~\IEEEmembership{Member,~IEEE}, Mohamed-Slim Alouini,~\IEEEmembership{Fellow,~IEEE}, Tareq Y. Al-Naffouri,~\IEEEmembership{Senior Member,~IEEE}
\thanks{This work is supported by the Office of Sponsored Research (OSR) at King
Abdullah University of Science and Technology (KAUST).

The authors are with the Department of Electrical Engineering, Computer Electrical and Mathematical Sciences \& Engineering (CEMSE) Division, King Abdullah University of Science and Technology (KAUST), Thuwal, Makkah Province, Kingdom of Saudi Arabia, 23955-6900. This paper is an extension of our previous work in \cite{Nasir2018icc}.}
}
\IEEEtitleabstractindextext{
\begin{abstract} \justifying
Underwater optical wireless links have limited range and intermittent connectivity due to the hostile aquatic channel impairments and misalignment between the optical transceivers. Therefore, multi-hop communication can expand the communication range, enhance network connectivity, and provide a more precise network localization scheme. In this regard, this paper investigates the connectivity of underwater optical wireless sensor networks (UOWSNs) and its impacts on the network localization performance.  Firstly, we model UOWSNs as randomly scaled sector graphs where the connection between sensors is established by point-to-point directed links. Thereafter, the probability of network connectivity is analytically derived as a function of network density, communication range, and optical transmitters' divergence angle. Secondly, the network localization problem is formulated as an unconstrained optimization problem and solved using the conjugate gradient technique. Numerical results show that different network parameters such as the number of nodes, divergence angle, and transmission range significantly influence the probability of a connected network. Furthermore, the performance of the proposed localization technique is compared to well-known network localization schemes and the results show that the localization accuracy of the proposed technique outperforms the literature in terms of network connectivity, ranging error, and number of anchors.  
  \end{abstract}

}

\maketitle
\section{Introduction}
Underwater wireless sensor networks (UWSNs) are the enabler of many underwater observation systems which span a wide range of applications including instrument monitoring, climate recording,  prediction of natural disasters, exploration for the oil industry, search \& rescue missions, and marine life study \cite{saeed2018underwater}. UWSNs also play a key role to control autonomous underwater vehicles (AUVs), stand-alone applications, and supervision of the cabled underwater communication systems which deploy an extensive amount of sensor nodes (seismometers, wave sensors, cameras, etc.) over miles of coverage on the ocean floor \cite{Tunnicliffe2008}. 

A significant part of the electromagnetic frequency spectrum suffers from unique underwater wireless channel conditions and display severe attenuation characteristics, frequency dispersion, and multipath fading \cite{Kaushal2016underwater}. In the past decades, acoustic systems have therefore received considerable attention thanks to their long communication ranges. Nonetheless, underwater acoustic communication has low achievable rates (10-100 kbps) due to their limited bandwidth and low propagation speed (1500 m/s) \cite{7593257}. Fortunately, these shortcomings of acoustic systems can be augmented by underwater optical wireless communication (UOWC) which has the advantage of higher achievable rates, lower latency, and enhanced security \cite{Kaushal2016underwater}. However, UWOC has a very limited range attainability (10-100 m) as a result of aquatic channel impairments (i.e., scattering, absorption, and oceanic turbulence, etc.) and noise sources such as sunlight, background, thermal, and dark current noises \cite{Akhoundi2016cellular, Oubei15}. 

Range limitation of UWOC can be augmented with multi-hop UOWSNs where nodes can share information for long distances through intermediate nodes \cite{Celik2018modeling}. Indeed, multi-hop cooperative communications have been extensively studied for RF networks \cite{Uysal2009}, acoustic underwater networks \cite{Liao2016}, and terrestrial wireless optical networks \cite{alquwsiee2015}.  Due to the omnidirectional communication capability of RF and acoustic signals, wireless sensor networks are traditionally modeled as geometric random graphs \cite{Gupta1998} where two sensor nodes $n_i$ and $n_j$ are generally assumed to establish a bidirectional communication link (i.e., $n_i \leftrightarrows n_j$). On the contrary, such a model is not suitable for UOWSNs because a node can only reach to the nodes within a certain beam scanning angle around their transmission trajectory, that is, optical wireless nodes are connected via unidirectional links. Directed communication networks are generally modeled by random scaled sector graphs \cite{Wu2014} where a unidirectional communication link from node $n_i$ to $n_j$ (i.e., $n_i \rightarrow n_j$) is established if and only if $n_j$ is positioned within the beam scanning angle of $n_i$. Notice that a directed reverse path is possible (i.e., $n_j \rightarrow n_i$) if  $n_i$ is in the beam-width of $n_j$ or through other multi-hop path.

There exists a reciprocal relationship between the degree of network connectivity and the performance of localization, with each susceptible to be influenced by the other. Connectivity of a network is often used as a metric for different performance parameters such as survivability, robustness, and fault tolerance \cite{Dall2002}. It is measured by a number of links in the network and a network is referred to be as connected if there exists at least one connecting path between any two nodes in the network. In this regard, the connectivity is also closely related to the link reliability that is mainly affected by the pointing errors and misalignment of the optical transceivers. Pointing and alignment errors are generally caused by random movements of the sea surface \cite{Dong2013, Zhang2015}, depth depended variations and deep currents \cite{Johnson2013}, and oceanic turbulence \cite{Yi2015}. Therefore, there is a dire need for a accurate pointing, acquisition, and tracking (PAT) mechanisms of optical transceivers to sustain reliable single-hop links. That being said, precise localization of UOWSN is of utmost importance because of three reasons: 1) PAT mechanisms can work properly only if the location of the target node is known with a certain accuracy, 2) Effective geographical routing schemes can be developed for multi-hop communications, and 3) Gathered data is useful only if it refers to a particular position of the sensor node. Network localization is especially useful for a number of applications such as target detection, intruder detection, routing protocols, and data tagging. On one hand, a better network connectivity substantially enhances the localization performance since having more pairwise range measurements intuitively reduces the localization errors. On the other hand, a more accurate location information yields a more precise PAT mechanism to sustain reliable single-hop links. Furthermore, network connectivity and localization can also be considerably enhanced by multi-hop communication over these reliable single-hop links, which can be enabled by effective geographic routing algorithms relying upon the precise node locations.

\begin{figure}[t]
\begin{center}  
\includegraphics[width=0.95 \columnwidth]{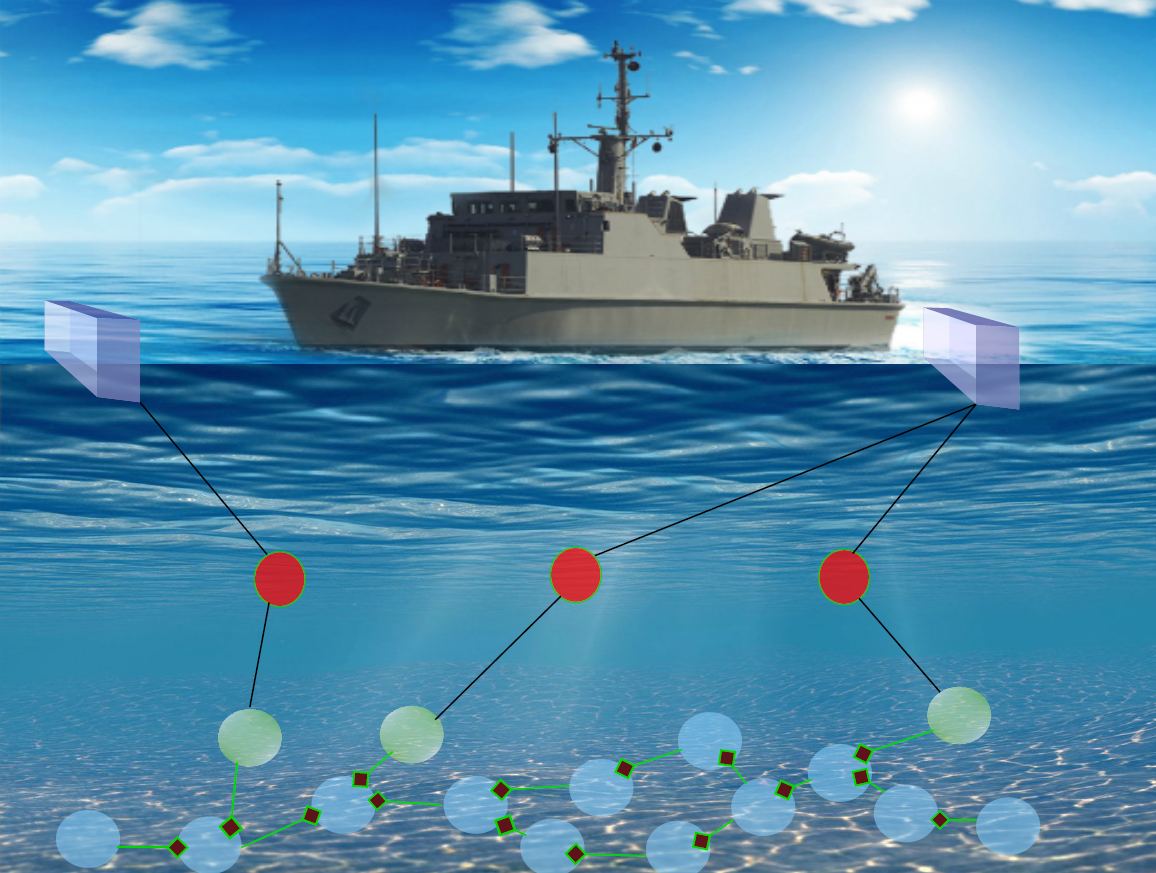}  
\caption{Multi-hop underwater optical sensor network setup.\label{fig:1}}  
\end{center}  
\vspace{-1.5 em}
\end{figure}

Existing research efforts on network connectivity and localization can be exemplified as follows:
The problem of network connectivity is addressed in \cite{Betstetter2004} for omnidirectional networks such that no node is obscured by RF wireless sensor networks. In \cite{Ferrero2014}, the authors have analyzed the coverage probability of random sector graphs for RF wireless sensor networks. In \cite{Diaz2003, Okorafor2009}, the authors have studied the circuit based routing paradigm for node isolation property of terrestrial optical wireless networks where the network connectivity goal cannot be achieved in the presence of an obscured node. A connectivity framework for UOWSNs is discussed in \cite{cowsn} where the authors assume bidirectional links between every pair of sensor nodes, which is not practical in UOWSNs. A number of acoustic underwater sensor networks localization techniques have been proposed in the past \cite{1acoustic,2acoustic, 3acoustic,  4acoustic, 5acoustic, 6acoustic, kantarci2011survey}. The performance of every localization technique mainly relies on the initial reference position, number of sensor nodes, ranging technique, number of anchors, and the position of the anchors in the network \cite{TAN20111663}. However, the aforementioned optical communication challenges do not allow the use of existing acoustic localization techniques for underwater optical sensor nodes. A UOWSN localization method is proposed in \cite{Akhoundi2017underwater} where received signal strength (RSS) and time of arrival (ToA) methods are investigated for an optical code-division multiple access networks.  However, authors assume omnidirectional optical wireless communication which is not always available for optical wireless communication. In \cite{Saeed2017, Nasir2018limited, Nasir2018twc}, RSS based localization schemes for UOWSNs have been proposed where each optical sensor node scans a circular region for ranging and the links between the nodes are assumed to be bi-directional.  In \cite{Saeed2018tcom}, the authors have proposed an RSS based localization scheme for UOWNs which takes into account outliers and optimizes the anchor positions. Nevertheless, \cite{Saeed2017, Nasir2018limited, Nasir2018twc, Saeed2018tcom} do not take the connectivity analysis into account which is important for network localization schemes. Whereas, in this paper, the optical sensor nodes are able to communicate only within its angular sector and the links are uni-directional. Due to the aforementioned constraints of connectivity and directional angular links, the localization problem of a UOWSN becomes more practical but also a challenging task.

\subsection{Main Contributions}
Our main contributions can be summarized as follows: 
\begin{itemize}
\item 
A stochastic network connectivity analysis is developed based on the network parameters of multi-hop UOWSNs. Unlike the symmetrical bi-directional graphs of traditional sensor networks, we model UOWSNs as uni-directional random graphs where coverage region of an optical transmitter node is angular-sector shaped. Accordingly, we define descendant and antecedent neighbors and derive a closed-form expression of the probability of $K$-connectivity for multi-hop UOWSNs as a function of beam scanning angles, transmission range, and the number of nodes. Even the derived expressions are general enough to be applicable for any directional and asymmetric graphs (including free-space optical communication networks), we especially focused on multi-hop UOWSNs since channel impacts on the number of nodes, beam scanning angles, and transmission range shows the benefit of multi-hop communications more significantly.  

\item 
The localization of multi-hop UOWSNs is formulated as an unconstrained optimization problem and solved using the conjugate gradient technique. We also evaluate the impact of connectivity, ranging error, and the number of anchors on the performance of the proposed localization technique. 

\item 
Finally, analytic findings are verified with extensive simulation results for different system parameters such as the number of nodes, transmission range, and beam scanning angles. Localization performance of the proposed method is also compared with other well-known network localization schemes such as multidimensional scaling (MDS) \cite{Shang, nasir2015, Rajawat2017}, and distance vector routing (DV) \cite{Niculescu.D}.  
\end{itemize}

\subsection{Paper Organization}
The rest of the paper is organized as follows: In Section  \ref{sec:relatedwork} brief overview of the related work is presented. Section \ref{sec:networkmodel} introduces the system model for a multi-hop UOWSN. Section \ref{sec:analysis} and \ref{sec:localization} present the stochastic analysis of network connectivity and elaborate the design of the proposed localization system, respectively. In Section \ref{sec:results}, simulations are conducted for the performance evaluation of the proposed localization system. We conclude the paper in Section \ref{sec:conc}.  

\section{Related Work}\label{sec:relatedwork}

In UOWSNs, cooperation and connectivity between the sensor nodes are important factors for maintenance and network survivability. The authors in \cite{Ammari2008} discussed several issues related to the connectivity of the omnidirectional sensing networks and proposed different assessment models. In \cite{Ammari2006}, the authors have investigated the network coverage problem for heterogeneous and homogenous omnidirectional wireless sensor networks. The probability of isolated nodes has been investigated in \cite{Tseng2010} where the mobile users visit the isolated nodes and collect the sensing data. In \cite {Dini2008}, the authors have investigated the connectivity of omnidirectional wireless sensor networks and proposed a method to stitch the network in case there are isolated nodes. In \cite{Diaz2003, Okorafor2009}, the authors have studied the circuit based routing paradigm for node isolation property of terrestrial optical wireless networks where the network connectivity goal cannot be achieved in the presence of an obscured node. Connectivity analysis of UOWNs have been performed in \cite{cowsn} where the authors have assumed omnidirectional links between the nodes which do not hold in case of directive optical wireless communications.

Besides the problem of connectivity, localization of nodes in UOWNs is of great importance which can enable numerous applications. Conventionally, localization methods are either range based or range free where the range based methods rely on different ranging methods to estimate the distances, and then, estimate position of the node based on the estimated distances. The range-free localization methods provide coarse position estimation where usually the area containing the node is estimated. Since range-free localization methods are not yet developed for UOWNs, we focus on range based methods.

The range-based localization methods for UOWNs can be classified into two categories as distributed and centralized methods. The authors in \cite{Akhoundi2017underwater} have proposed for the first time an RSS and ToA based distributed localization method. The authors have considered an optical base station (OBS) placed in a hexagonal cell which serves as an anchor for the users. A hybrid acoustic and optical RSS ranging based localization method have been proposed in \cite{Saeed2017} where the authors have considered a weighting strategy to give more importance to accurate ranging measurements. The users are able to estimate the distance to multiple OBSs and then estimate its position by using linear least square estimation. A centralized RSS based localization method has been proposed in \cite{Nasir2018limited}, where the nodes estimate and forward the single hop RSS based distances to the centralized node. The centralized node is then able to estimate the position of each node. 

In the past decade, connectivity and coverage issues of underwater acoustic wireless sensor networks (UAWSNs) have been well studied. In \cite{ SENEL2015}, the authors have proposed a connected dominating set based strategy to improve the coverage and connectivity for UAWSNs where the connectivity and coverage improve with the increase in density and transmission range of the sensor nodes. Consequently, in \cite{ Han2015}, the authors have analyzed the impact of deployment on localization performance for UAWSNs where the authors have numerically shown that tetrahedron deployment scheme has better localization performance as compared to the random deployment. Furthermore, a localization scheme which takes advantage of the sensors mobility have been presented in \cite{ Akbas2015} for UAWSNs. However, none of the existing literature studies the problem of connectivity and its impact on the localization for UOWNs.
Therefore, it is important to investigate the problem of connectivity and localization for UOWNs due to their unique behavior. The hostile underwater environment makes it very challenging to develop a connected network and find the location of each node. The dynamic and temporal changes in the behavior of the underwater optical wireless channel greatly influence the communication range and beam widths of the transmitted signals. Additionally, localization for multihop communication requires to develop robust matrix completion strategies to mitigate the shortest path estimation ranging error. Therefore, we propose a novel matrix completion strategy to reduce the shortest path estimation error and then evaluate the performance of the proposed localization method in terms of connectivity, ranging error and the number of anchors in the network. 
\begin{figure}[t]
\begin{center}  
\includegraphics[width=0.95 \columnwidth]{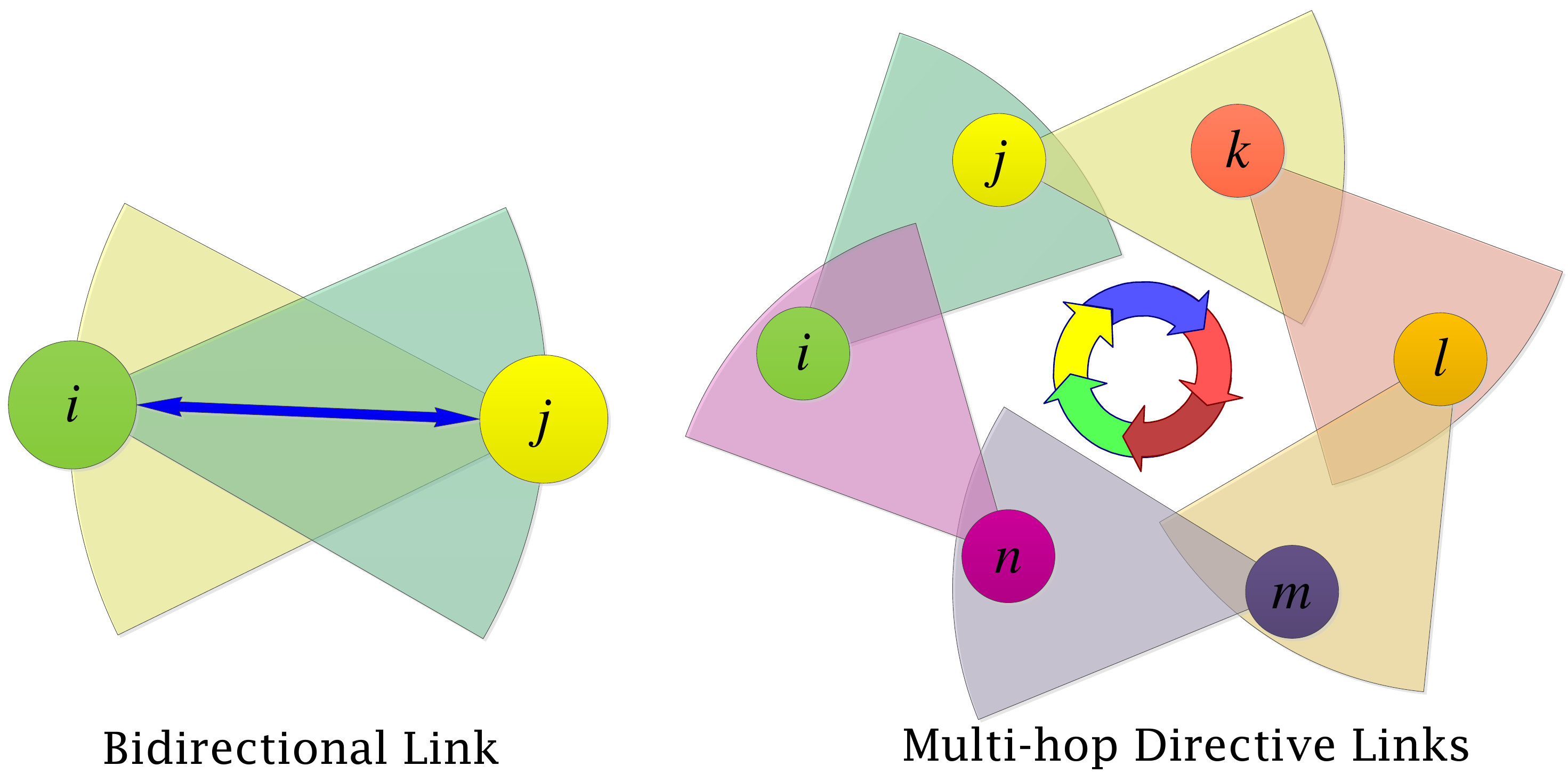}  
\caption{Connection types in random sector directed graphs.
\label{fig:2}}  
\end{center}  
\end{figure}
\begin{figure}[t]
\begin{center}  
\includegraphics[width=0.95 \columnwidth]{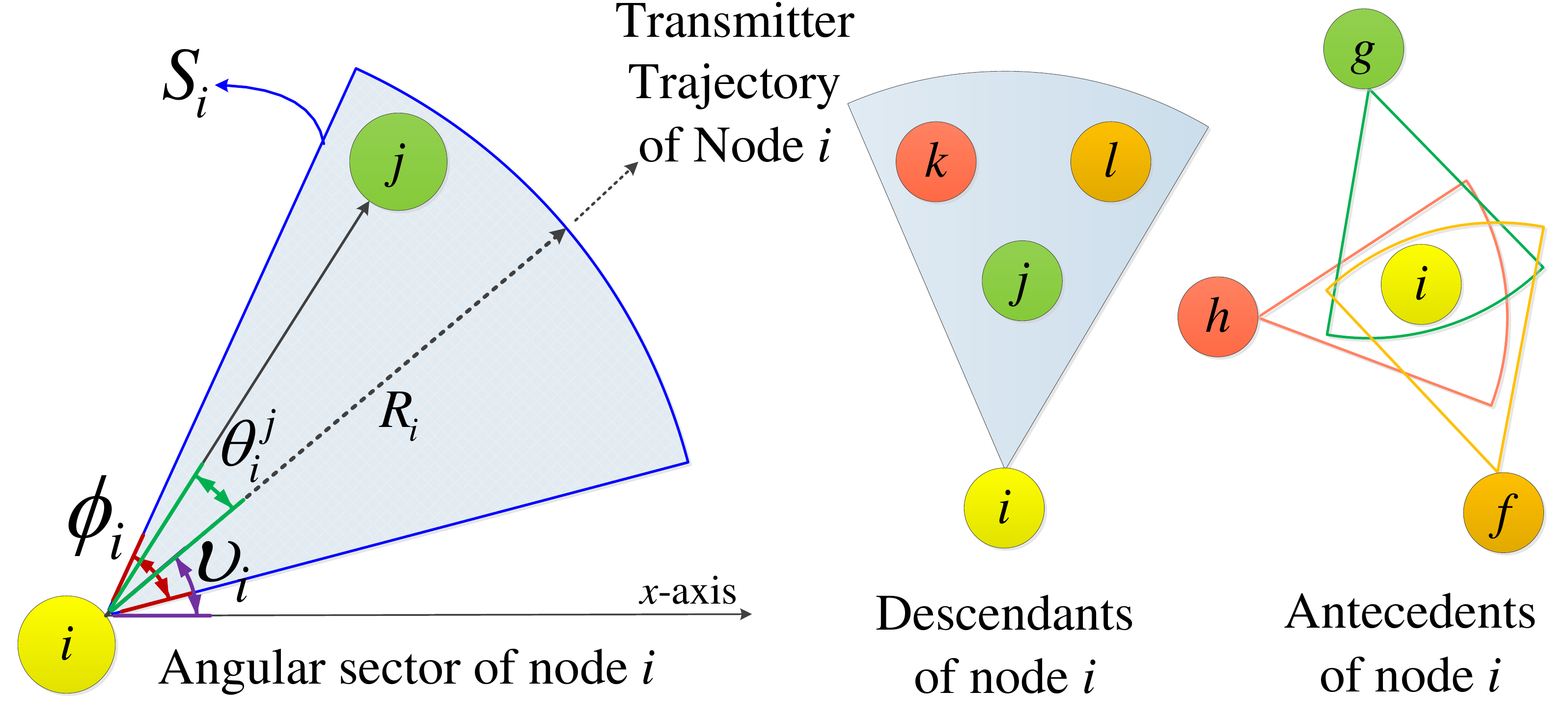}  
\caption{Demonstration of connectivity parameters, descendants, and antecedents of $n_i$.\label{fig:p}}  
\end{center}  
\vspace{-1.5 em}
\end{figure}

\section{System Model}
\label{sec:networkmodel}
In this section, we first model UOWSNs as random sector graphs and then present considered underwater optical wireless channel model.
\subsection{Network Model}
We assume that the spatial distribution of optical wireless nodes follows a homogeneous Poisson point process (PPP) within a finite area of $A$ as shown in Fig.~\ref{fig:1} where the blue circles represent the optical sensor nodes with narrow beam scanning angles, embedded on the seafloor. Green optical sensor nodes are floating nodes with wider beam scanning angles\footnote{{\color{black} We assumed that floating relay nodes are fastened to the seabed in order to prevent their loss by drifting in case of strong deep currents.}} and red circles are the anchor nodes. The anchor nodes are able to communicate with the floating nodes and the surface buoys, represented by the rectangles. Finally, the surface buoys communicate through RF medium with the surface station or the ship to transmit the sensing data. Since coverage area of an optical wireless node is characterized by the divergence angle of the light beam and thus angular sector-shaped, we consider two type of links: bi-directional and multi-hop directive links. As illustrated in Fig.~\ref{fig:2}, node $n_i$ is connected to $n_j$ ($n_i \rightarrow n_j$) if $\mathbf{c}_i$ (coordinates of sensor node $n_i$) falls within the angular sector ($S_j$) of node $n_j$, and vice versa. Thus, a bidirectional link between $n_i$ and $n_j$ ($n_i \leftrightarrow n_j$) exists only if $n_i \rightarrow n_j$ and $n_j \rightarrow n_i$. Fig.~\ref{fig:2} also shows a scenario of multi-hop directive links with six optical sensor nodes, where $n_i$ communicates directly with $n_j$ whereas $n_j$ can communicate with $n_i$ over a multi-hop path through $n_k \rightarrow n_l \rightarrow n_m \rightarrow n_n \rightarrow n_i$. Accordingly, UOWSNs can be defined as a random directed graph as follows:

\newtheorem{definition}{Definition} 
\begin{definition}[UOWSNs as Random Sector Directed Graphs] \it
Denoting the total number of optical nodes by $M$, the scanning sector (coverage area) of $n_i, 1\leq i \leq M$, is defined as a tuple of random orientation $\zeta_i$, scanning angle $\phi_i$, communication range $R_i$, and sensor node coordinates $\mathbf{c}_i$, i.e., $\mathbf{S}_i=(\zeta_i,\phi_i, R_i, \mathbf{c}_i)$ which is illustrated in Fig. \ref{fig:p}. Accordingly, UOWSNs can be defined as a random sector directed graph $\mathcal{G}(\mathbf{\mathcal{V}},\mathbf{\mathcal{E}})$  where $\mathbf{\mathcal{V}} = \{ \mathbf{c}_1, \ldots, \mathbf{c}_i, \ldots, \mathbf{c}_M \}$ represent the set of vertices and $\mathbf{\mathcal{E}} \in \{0,1\}^M$ is the set of links which is primarily characterized $\mathbf{S}=\mathbf{S}_1, \ldots, \mathbf{S}_i, \ldots, \mathbf{S}_M$. Notice that $\mathbf{\mathcal{E}}_{i,j}=1$ only if $n_i \rightarrow n_j$ holds.
\end{definition} 

\begin{figure*}[!htb]
\centering
\minipage{0.48 \textwidth}
\centering
    \includegraphics[width=0.95 \columnwidth]{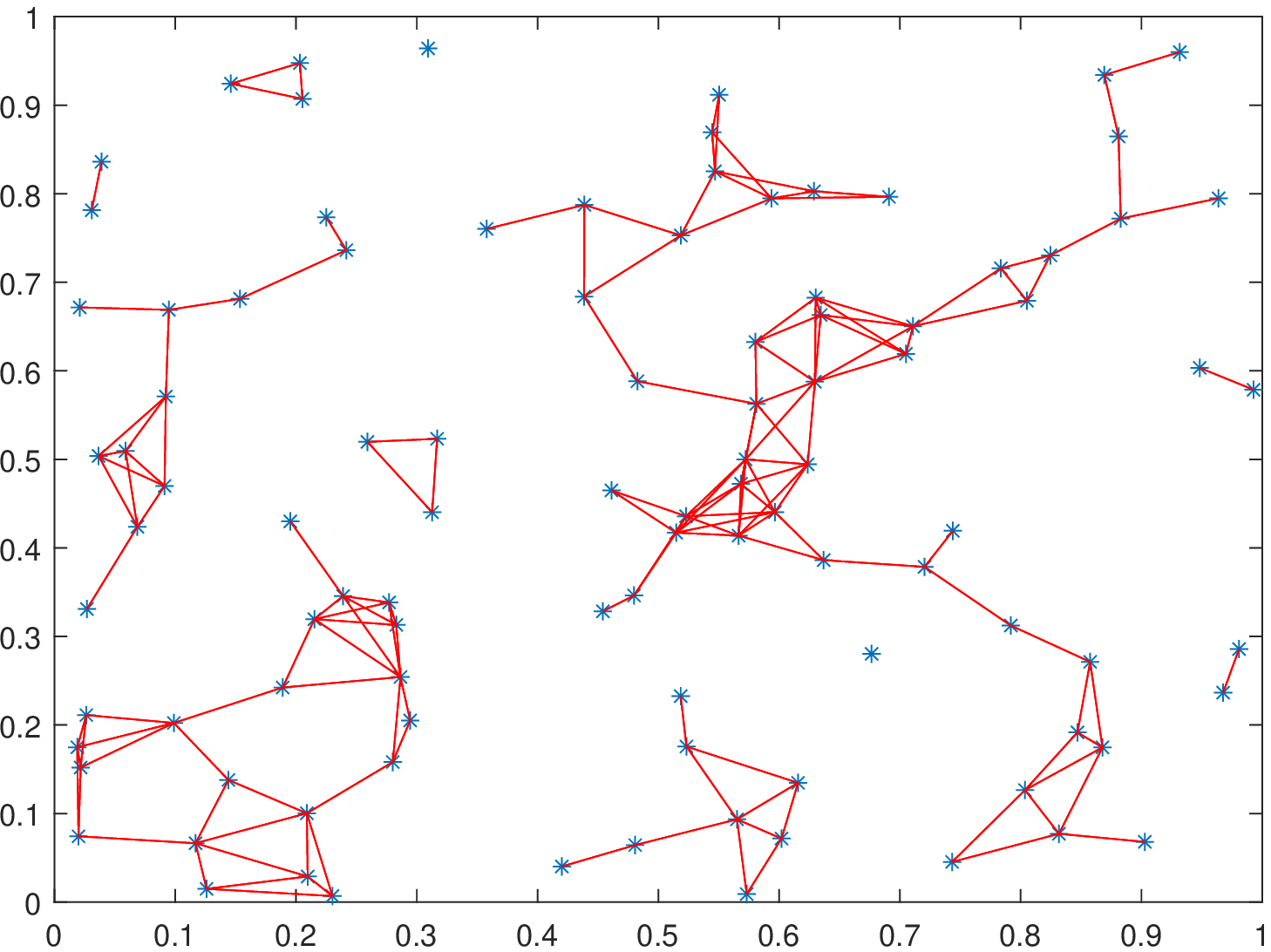}
    \caption{Random directed graph for $M=100$, $\phi = \frac{\pi}{3}$, and $r=0.2 m$.}
    \label{fig:pibythree}
\endminipage 
\minipage{0.48 \textwidth}
\centering
      \includegraphics[width=0.95 \columnwidth]{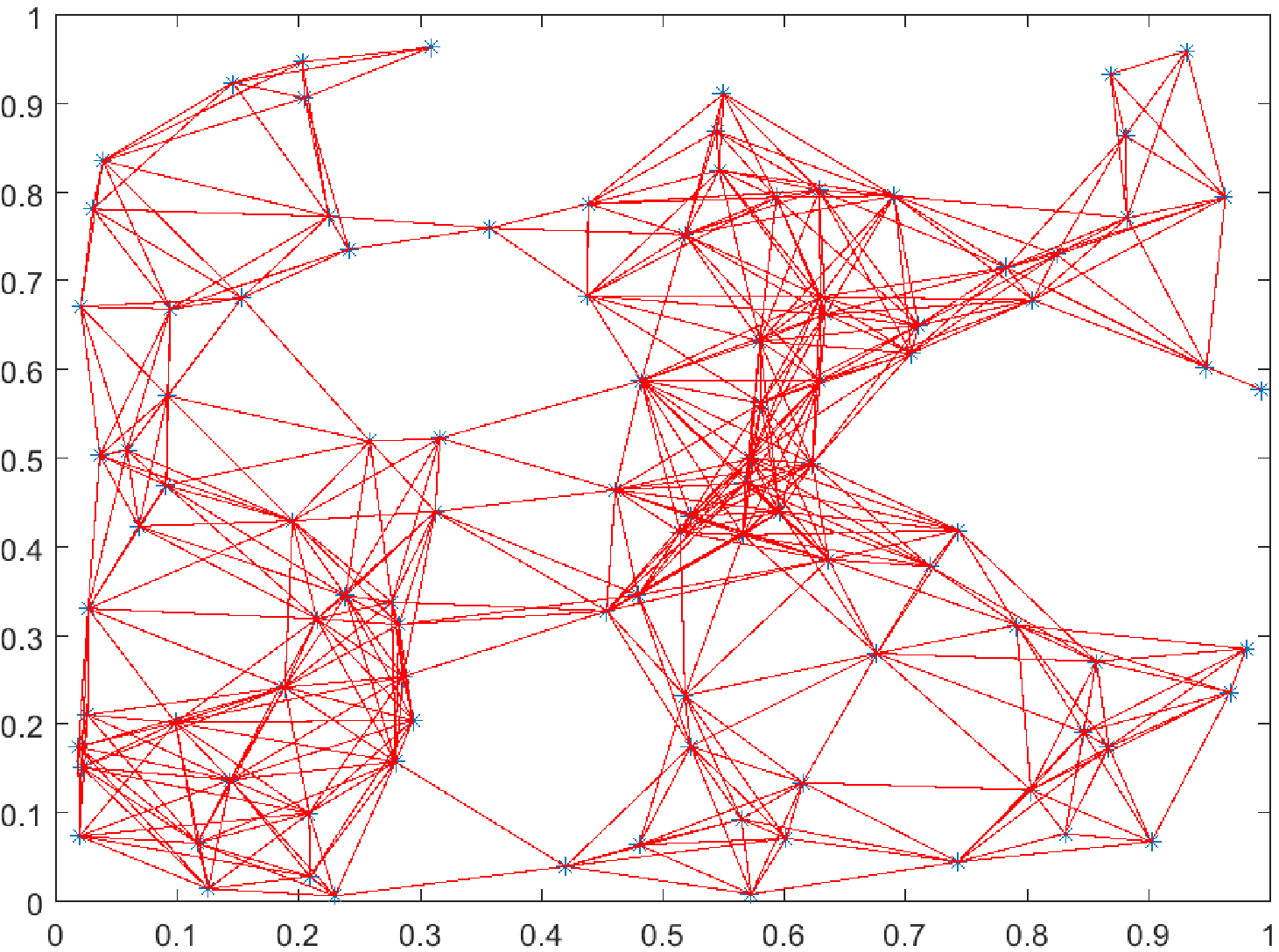}
      \caption{Random directed graph for $M=100$, $\phi = 2\pi$, and $r=0.2 m$. }
      \label{fig:twopi}
\endminipage 
\vspace{5pt}
\hrule
\end{figure*}

Random sector directed graphs and random geometric graphs are identical in case of $\phi = 2\pi$ \cite{Erdos1960, Martin1997, Dall2002, Wu2014}. Notice that two nodes $i$ and $j$ are connected when the distance between them is less than $R$ in random geometric graphs, however, the connectivity of random directed sector graphs also depends on the beam scanning angle and its orientation. Fig.~\ref{fig:pibythree} and Fig.~\ref{fig:twopi} show two different random directed sector graphs with scanning angles of $\phi = \frac{\pi}{3}$ and $\phi = 2\pi$, respectively. It is obvious that increasing the scanning angle for each node from $\phi = \frac{\pi}{3}$ to $\phi = 2\pi$, increases the number of links in the graph. These asymmetric and directional characteristics of the random directed sector graphs require us to define descendant and antecedent neighbors for every node.  
\begin{definition}[Descendant and Antecedent Nodes] \it
While descendants of node $n_i$ are defined as $\mathcal{D}_i \triangleq \{n_j | \: ~\forall~j:~\mathcal{E}_{i,j}=1\}$, i.e., the set of nodes who lies within the coverage region of $n_i$, antecedents of $n_i$ are defined as $\mathcal{A}_i \triangleq \{n_j | ~\forall~j:~\mathcal{E}_{j,i}=1 \}$  the set of nodes who can reach to $n_i$. 
\end{definition}
In Fig.~\ref{fig:p}, the set of descendants and antecedents of $n_i$ are shown as $\{n_j,n_k,n_l\}$ and $\{n_g,n_h,n_f\}$, respectively. As descendants and antecedents of a node may differ in practice from traditional omnidirectional wireless sensor networks, these inherent features lead us to a consider a distinct stochastic connectivity analysis, which is addressed in \ref{sec:analysis}. It is worth noting that proposed solutions are also applicable to terrestrial optical wireless networks (TOWNs) because of shared features of the directivity and sector-shaped coverage region, underwater channel impediments and their impacts on the range, connectivity, and localization performance is still different from TOWNs and worth investigation.

\subsection{Underwater Optical Wireless Channel Model}
\begin{figure}
\centering
\includegraphics[width=0.99\columnwidth]{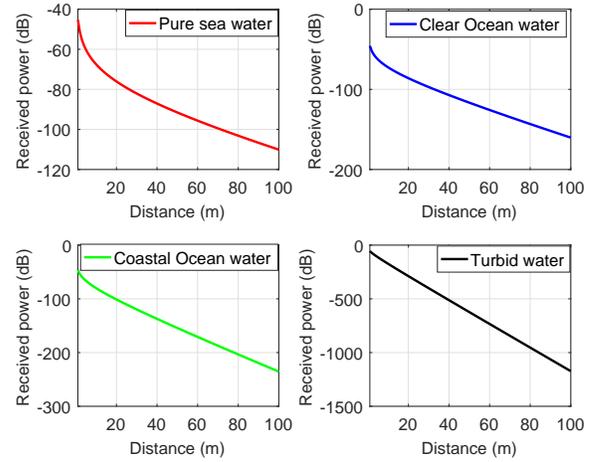}  
\caption{Received power (in dB) Vs. distance (m) for different types of water.\label{fig:watertype}}  
\end{figure}

The underwater aquatic medium consists of different elements with different concentrations which are either suspended or dissolved in the pure water \cite{Haltrin:99}. Due to these elements, emitted light suffers from absorption and scattering effects during the propagation within the aquatic medium.  In this paper, we consider Haltran's model which models the extinction coefficient $e(\lambda)$ as a combination of absorption coefficient $b(\lambda)$ and scattering coefficient $s(\lambda)$ as follows
\begin{equation}
e(\lambda) = b(\lambda) + s(\lambda),
\end{equation}
where $\lambda$ is the operating wavelength. The underwater chlorophyll is considered to be the major cause of light absorption in the wide wavelength ranges. The absorption coefficient $b(\lambda)$ is therefore expressed as
\begin{equation}
b(\lambda) = b_w(\lambda)+b_{cl}(\lambda)+b_f C_f \exp^{-\kappa_{f}\lambda}+b_h C_h \exp^{-\kappa_{h}\lambda},
\end{equation}
where $\kappa_{f}$ and $\kappa_{h}$ are the constants, $ b_w(\lambda)$ represents the pure water absorption, $b_{cl}(\lambda)$ is the chlorophyll absorption coefficient, $b_f = 35.959~m^2/mg$ is the absorption coefficient of fulvic acid, $b_h = 18.828~m^2/mg$ is the absorption coefficient of humic acid, $C_f$ is the concentrations of fulvic acid, and $C_h$ represents the concentrations of humic acid. $C_f$ and $C_h$ are given as \cite{Haltrin:99}
\begin{equation}
C_f = 1.74098 C_e \exp^{(0.12327\frac{C_e}{C^0_e})},
\end{equation}
and
\begin{equation}
C_h = 0.19334 C_e \exp^{(0.12343\frac{C_e}{C^0_e})},
\end{equation}
where $0 \leq C_e \leq 12~mg/m^2$ and $C^0_e = 1 ~ mg/m^3$. Similarly the scattering coefficient $s(\lambda)$ is modeled as
\begin{equation} \label{eq: scattering}
s(\lambda) = s_{\omega} + s_s^0(\lambda)C_s + s_l^0(\lambda)C_l,
\end{equation}
where $s_{\omega}$ is the scattering coefficient for pure water, $s_s^0(\lambda)$ is the scattering coefficient for small particles, $s_l^0(\lambda)$ is the scattering coefficient for large particles, $C_s$ is the concentration of small particles, and $C_l$ represents the concentration of large particles. All parameters in \eqref{eq: scattering} are defined as \cite{Haltrin:99}
\begin{align}
s_{\omega} &= 0.005826 \left(\frac{400}{\lambda}\right)^{4.322},\\
s_s^0(\lambda) &= 1.151302 \left(\frac{400}{\lambda}\right)^{1.7},\\
s_l^0(\lambda) &= 0.341074 \left(\frac{400}{\lambda}\right)^{0.3}, \\
C_s &= 0.01739 C_e \exp{\left\{0.11631\frac{C_e}{C^0_e}\right\}}, \text{ and} \\
C_l &= 0.76284 C_e \exp{\left\{0.03092\frac{C_e}{C^0_e}\right\}}.
\end{align}
Contingent upon the extinction coefficient and hardware specifications, transmitted signal power from node $n_i$ received at node $n_j$ is based on the following link budget formula \cite{Arnon:09, Gkoura2017} 
\begin{equation}
\label{eq:Pr}
P_{i,j}^r = P_t^i \delta_t^i \delta_r^j \exp^{\left(\frac{-e(\lambda) d_{ij}}{\cos \theta_i^j}\right)}\frac{B_j^r \cos \theta_i^j}{2\pi R_{ij}^2(1- \cos \theta_0^i)}
\end{equation}
where $P_t^i$ is the transmission power of node $n_i$, $\delta_t^i$ ($\delta_r^j$) is the transmitter (receiver) optical efficiency of node $n_i$ (node $n_j$), $d_{ij}$ is the transmission range between nodes $n_i$ and $n_j$, $\theta_0^i$ is the transmitter's divergence angle, $B_j^r$ is the receiver aperture area, and $\theta_i^j$ is the angle between trajectory of the node $n_i$'s transmitter and node $n_j$'s receiver. Following from \eqref{eq:Pr},  the estimated range between the nodes $n_i$ and $n_j$ is given as
\begin{equation}
R_{ij} = \frac{2\cos\theta_{i}^j}{e(\lambda)}W_0\left(\frac{e(\lambda)}{2}\sqrt{\frac{P_t^i \delta_t^i \delta_r^j B_j^r \cos \theta_i^j}{P_{i,j}^r 2 \pi (1-\cos(\theta_0^i))}}\right)+\eta_i^j
\end{equation}
where $W_0$ is real part of Lambert $W$ function and $\eta_i^j$ is the ranging error modeled as zero mean Gaussian random variable {\color{black}$\eta_i^j~N(0, \sigma_{ij}^2)$} with variance $\sigma_{ij}^2$. Based on the simulation parameters in \cite{Arnon:09}, Fig.~\ref{fig:watertype} demonstrates the impact of absorption, scattering, and geometrical loss on the range estimation . It is clear from Fig.~\ref{fig:watertype} that increase in the turbidity of water (large extinction coefficient) results in low received power and thus yields limited transmission ranges. 

\section{Connectivity Analysis of UOWSNs}
\label{sec:analysis}
The number of descendants and antecedents nodes for $n_i$ are random variables denoted by $D_i\triangleq |\mathcal{D}_i|$ and $A_i\triangleq |\mathcal{A}_i|$, respectively. Defining the set of neighbors of $n_i$ as the union of its descendants and antecedents nodes (i.e., $\mathcal{K}_i= \mathcal{D}_i \cup \mathcal{A}_i$), a network is referred to as a $\mathrm{fw}_k$ forward-connected if $D_i \geq k, \forall _i$, otherwise it is $\mathrm{fw}_k$ forward-isolated/obscured where $k$ is the degree of connectivity. Similarly, a network is referred to as $\mathrm{bw}_k$ backward-connected if $A_i \geq k$, otherwise it is $\mathrm{bw}_k$ backward-isolated/obscured. As a special case, the network can be regarded as \textit{connected} if there is no obscured node in $\mathcal{G}(\mathcal{V},\mathcal{E})$, i.e., $D_i \geq 1$ and $A_i \geq 1, \forall i$.
\subsection{Stochastic Connectivity Analysis of UOWSNs}
In this section, we derive the probability of having a connected multi-hop UOWSN as a function of design parameters $M$, $R$ and $\phi$. Defining the probability of being forward-connected and backward-connected as $p^i_d = \text{Pr}[D_i \geq 1]$ and  $p^i_a = \text{Pr}[A_i \geq 1]$, respectively, probability of having $n_i$ non-obscured is given by \cite{Diaz2003}
\begin{equation}\label{eq: p0}
p^i_o =  p^i_d.p^i_{a|d},
\end{equation}
where $p^i_{a|d}$ is the probability that $n_i$ is backward connected given that it is also forward connected. Without loss of generality, let us consider a unit network area $A=1$ for the sake of simplicity. As $M\rightarrow \infty$ and $R<<1$, we have 
\begin{equation}\label{eq: descendent}
p^i_d = p^i_a= 1-\exp\{Q\}, \forall i,
\end{equation}
where $Q \triangleq {\frac{-\phi M  R^2}{2}}$ by assuming nodes have identical range and divergence angles. On the other hand, $p^i_{a|d}$ can be evaluated by considering the following two cases:
\begin{itemize}
\item Case 1: This is the case where $n_i$ does not have any bidirectional link with its descendant nodes, i.e.,  $D_i = 0$. As $p^i_{a|d}$ is conditioned on forward-connectivity, backward-connectivity of $n_i$ is guaranteed by a reverse path which consists of at most $M-1$ nodes.
\item Case 2: In this case, $n_i$ have at least one bi-directional link, i.e., at least one of the descendant node is also an antecedent node.
\end{itemize}
Both of these cases are disjoint and $p^i_{a|d}$ can be written as 
\begin{eqnarray}\label{eq: jointprob2}
\nonumber p^i_{a|d} &=& 1-\text{Pr}[A_i = 0 | D_i \geq 1] \\
&=& 1- \bigg(\text{Pr}[A_i = 0 | D_i \geq 1,~\Upsilon=0] \text{ Pr}[\Upsilon = 0] \bigg.\nonumber  \\
& & \bigg. + \text{ Pr}[A_i = 0 | D_i \geq 1, \Upsilon = 1] \text{ Pr}[\Upsilon = 1]\bigg),
\end{eqnarray}
where $\Upsilon$ is number of bidirectional links. One can observe from \eqref{eq: jointprob2} that the second term is contradictory, i.e., if $A_i = 0$ there are no bidirectional links, therefore $\text{Pr}[A_i = 0 | D_i \geq 1, \Upsilon = 1]=0$. Expression of $p^i_{a|d}$ is obtained in Appendix A as follows
\begin{eqnarray}\label{eq: jointprobfinal11}
p^i_{a|d}  &=& 1-\frac{\exp\{Q\}}{1-\exp\{Q\}}\left(1-\frac{\phi R^2}{2}\right)^{M-1} \times \nonumber \\ & &\left(\exp \left \{ \frac{-Q (2\pi - \phi)}{\pi(2-\phi R^2)} \right \} -1 \right), \forall i.
\end{eqnarray}
Substituting \eqref{eq: descendent} and \eqref{eq: jointprobfinal11} in \eqref{eq: p0} yields the following expression
\begin{eqnarray}\label{eq: jointprobfina}\nonumber
p_o &=& \left(1-\exp\{Q\}\right) \times \left[1-\frac{\exp\{Q\}}{1-\exp\{Q\}} \left(1-\frac{\phi R^2}{2}\right)^{M-1}\right.\\ && \left.\left(\exp \left \{ \frac{-Q (2\pi - \phi)}{\pi(2-\phi R^2)} \right \} -1 \right)\right]^M
\end{eqnarray}
It can be observed from \eqref{eq: jointprobfina} that as $\phi \rightarrow 2\pi$ and $M \rightarrow \infty$, $p_o \rightarrow1$ as it is expected..

\subsection{Evaluating the Probability of $k$ Connectivity}
In the previous section, we have derived the probability for a single obscured node or a connected network. Here, the expression for $k$ connectivity is derived for $k > 1$. From the previous discussion, node $i$ is $k$-connected with probability:
\begin{equation}
\label{eq:kk}
p^i_{o_{k}} =  p^i_{d_{k}}.p^i_{a_{k}|d_{k}},
\end{equation} 
where $ p^i_{d_{k}}$  is the probability of $k$-forward-connectivity, $p^i_{a_{k}}$ represents the probability of $k$-backward-connectivity and $p^i_{a_{k}|d_{k}}$ is the probability that $n_i$ is $k$-backward-connected given that it is also $k$-forward connected. The results from \eqref{eq: descendent} are extended for $k$-connectivity as
\begin{equation}\label{eq: pdk}
p^i_{d_{k}} = p^i_{a_{k}}= 1-\sum^{M-1}_{k>1}\frac{\exp\{Q\}(-Q)^k}{k !}, \forall i,
\end{equation}
and
\begin{equation}\label{eq: pdak}
p^i_{a_{k}|d_{k}} = \text{Pr}[A_i \geq k | D_i \geq k] = 1 - \text{Pr}[A_i < k | D_i \geq k].
\end{equation}
In similar to $k=1$ case, $p^i_{o_{k}}$ is derived from equation \eqref{eq: pdk} and \eqref{eq: pdak}. As an example here we derive the expression for $k=2$
\begin{equation}
p^i_{d_{2}} = \text{Pr}[D_i \geq 2] = 1-\text{Pr}[D_i = 0]-\text{Pr}[D_i = 1],
\end{equation}
where $\text{Pr}[D_i = 0] = \exp\{Q\}$ and $\text{Pr}[D_i = 1] = -Q\exp\{Q\}$. Therefore, 
\begin{equation}
p^i_{d_{2}} = p^i_{a_{2}}= 1-\exp\{Q\}\left(1-Q\right).
\end{equation}
The expression for $p^i_{a_{2}|d_{2}}$ is derived in appendix B.
Note that $p^i_{a_{2}|d_{2}}$ is equal to 1 when $\phi = 2\pi$. Probability of connectivity for higher degrees can be calculated following the similar steps in Appendix B.
\section{Localization  System for UOWSNs}
\label{sec:localization}
The knowledge of node coordinates is crucial in order to design a precise alignment algorithm and to associate the collected data with the sensing location as some of the observations are meaningful only with a precise location information.   
For a desirable localization performance, a better network connectivity is necessary because a well-connected network can provide more pairwise range measurements that intuitively reduces the localization errors. In order to realize reliable single-hop links, a more accurate location information should be leveraged to enable precise pointing, acquisition, and tracking (PAT) mechanisms of optical transceivers. Furthermore, network connectivity and localization can also be substantially improved by multi-hop communication over these reliable single-hop links, as already investigated in the previous section, which can be enabled by effective geographic routing algorithms based on the accurate node locations. In other words, there is a reciprocal relationship between the degree of network connectivity and performance of localization, with each susceptible to be influenced by the other.  
   However, limited connectivity of UOWSNs poses many challenges to acquire the accurate location information of the entire nodes in the network. In particular, all the distances to the surface station are not available due to the short transmission ranges, random orientation, and beam scanning angles of each node. Therefore, we propose a centralized localization system where the surface station collects the single hop neighborhood estimated distances and estimates the missing pairwise distances. Once the surface station computes the missing pairwise distances it estimates the position of all nodes by using at least three anchors for two-dimensional localization.
\subsection{Matrix Completion Strategy}
Let us consider that the observed distance matrix $\mathbf{\hat{D}} = \{\hat{d}_{i,j}\}_{i=1, i\neq j}^{M}$, where $\hat{d}_{ij} = d_{ij}+\eta_{ij}$ is the observed Euclidean distance, $d_{ij}$ is the actual Euclidean distance, and $\eta_{ij}$ is the ranging error between generic nodes $n_i$ and $n_j$.  The observation distance matrix at the surface station is given by
\begin{equation}\label{eq: distance}
\mathbf{\hat{D}} = \begin{bmatrix}
0 & \hat{d}_{12} & ? & \cdots & \hat{d}_{1M}\\
\hat{d}_{21} & 0 & ? & \cdots & ?\\
? & ? & 0 & \cdots & \hat{d}_{3M}\\
\vdots & \vdots & \vdots & \ddots & \vdots\\
\hat{d}_{M1} & ? & \hat{d}_{M3}& \cdots & 0
\end{bmatrix},
\end{equation}
\begin{figure*}
\centering
\minipage{0.475 \textwidth}
\includegraphics[width=1\columnwidth]{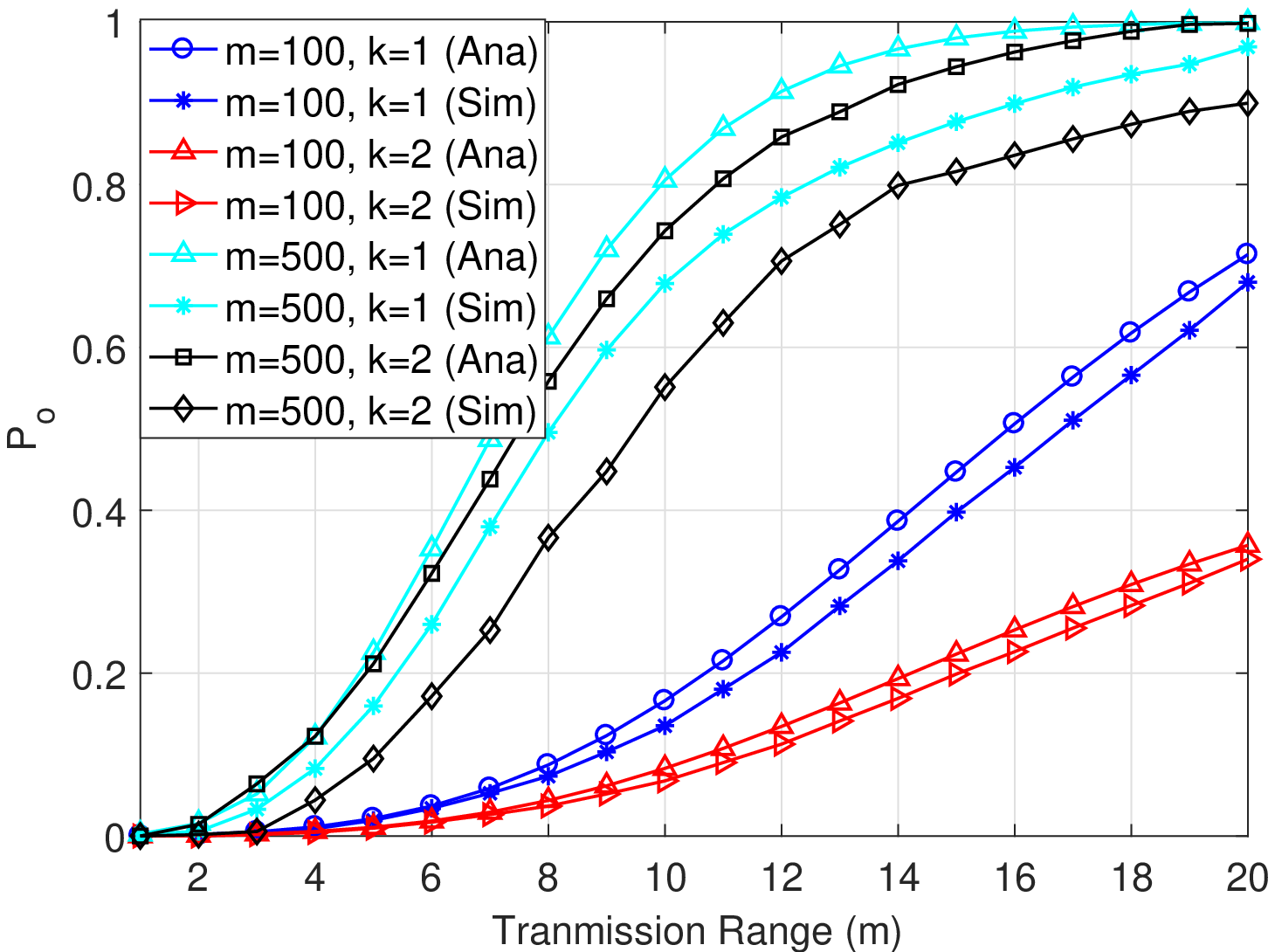}  
\caption{$p_o$ vs. transmission range for $\phi = \frac{2\pi}{9}$.\label{fig:4}}  
\endminipage
\minipage{0.475 \textwidth}
\includegraphics[width=1\columnwidth]{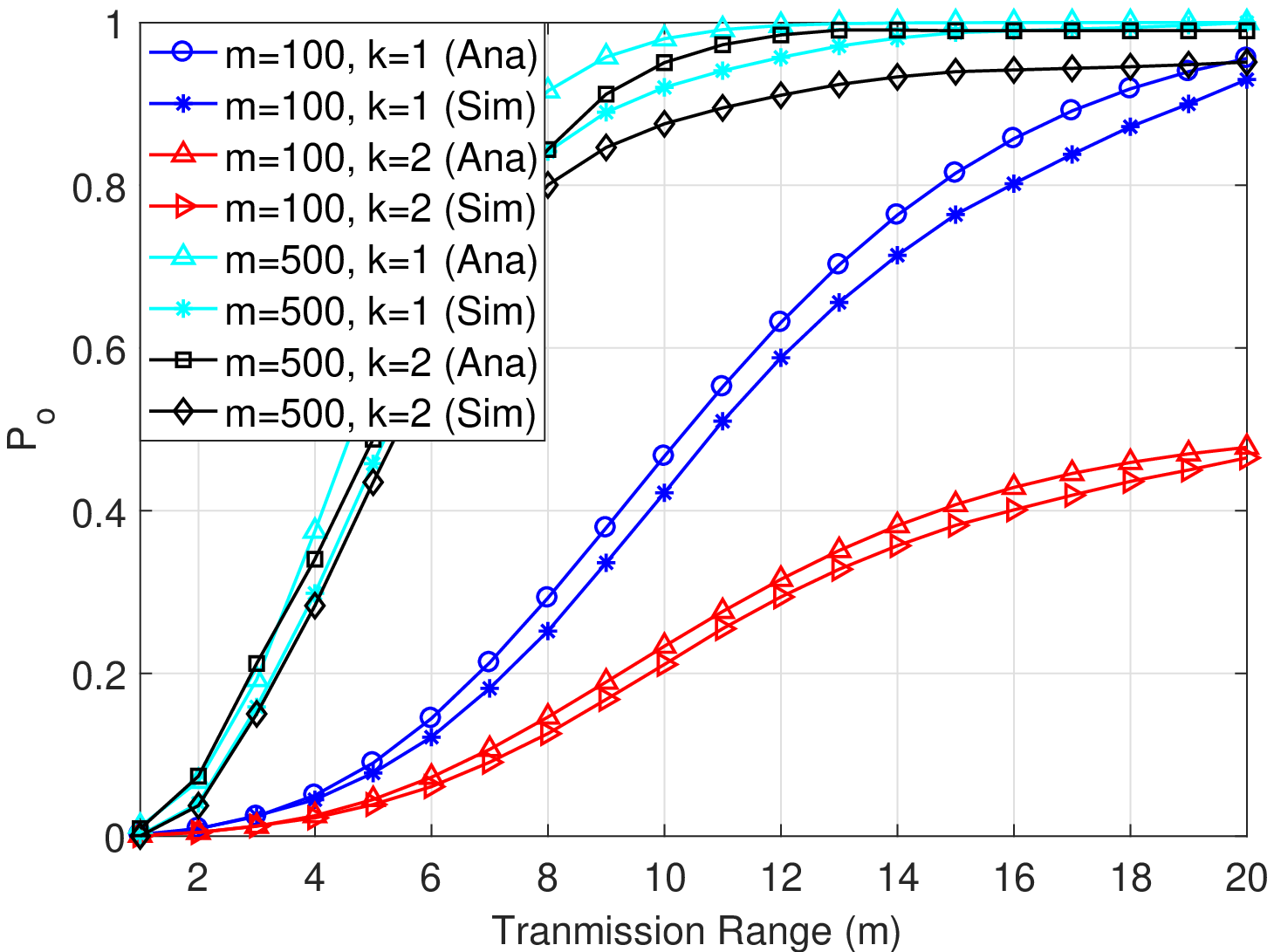}  
\caption{$p_o$ vs. transmission range for $\phi = \frac{\pi}{2}$.\label{fig:5}}  
\endminipage
\vspace{4pt}
\minipage{0.475 \textwidth}%
\includegraphics[width=1\columnwidth]{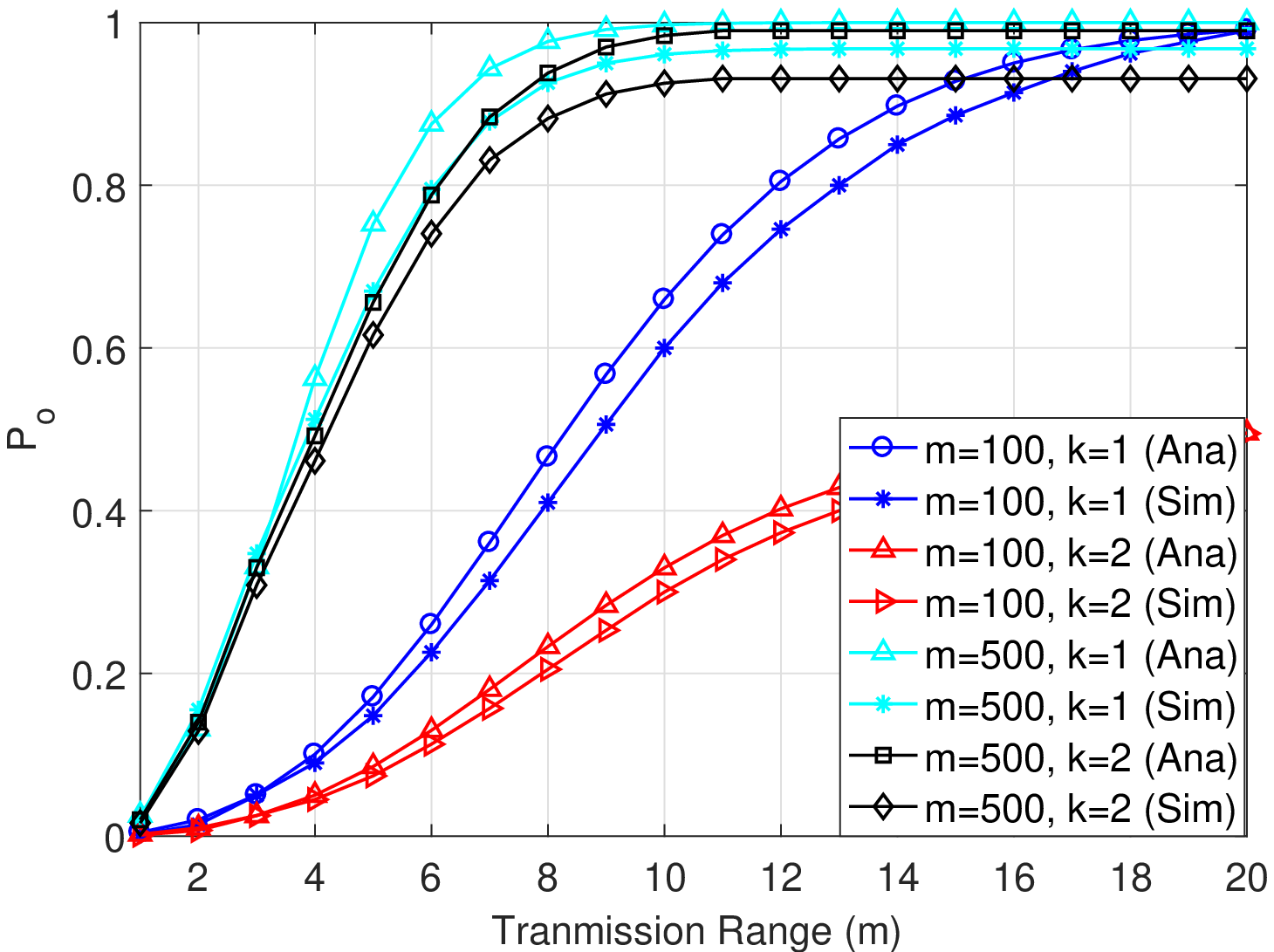}  
\caption{$p_o$ vs. transmission range for $\phi = \frac{3\pi}{4}$.\label{fig:6}}  
\endminipage
\minipage{0.475 \textwidth}
\includegraphics[width=1\columnwidth]{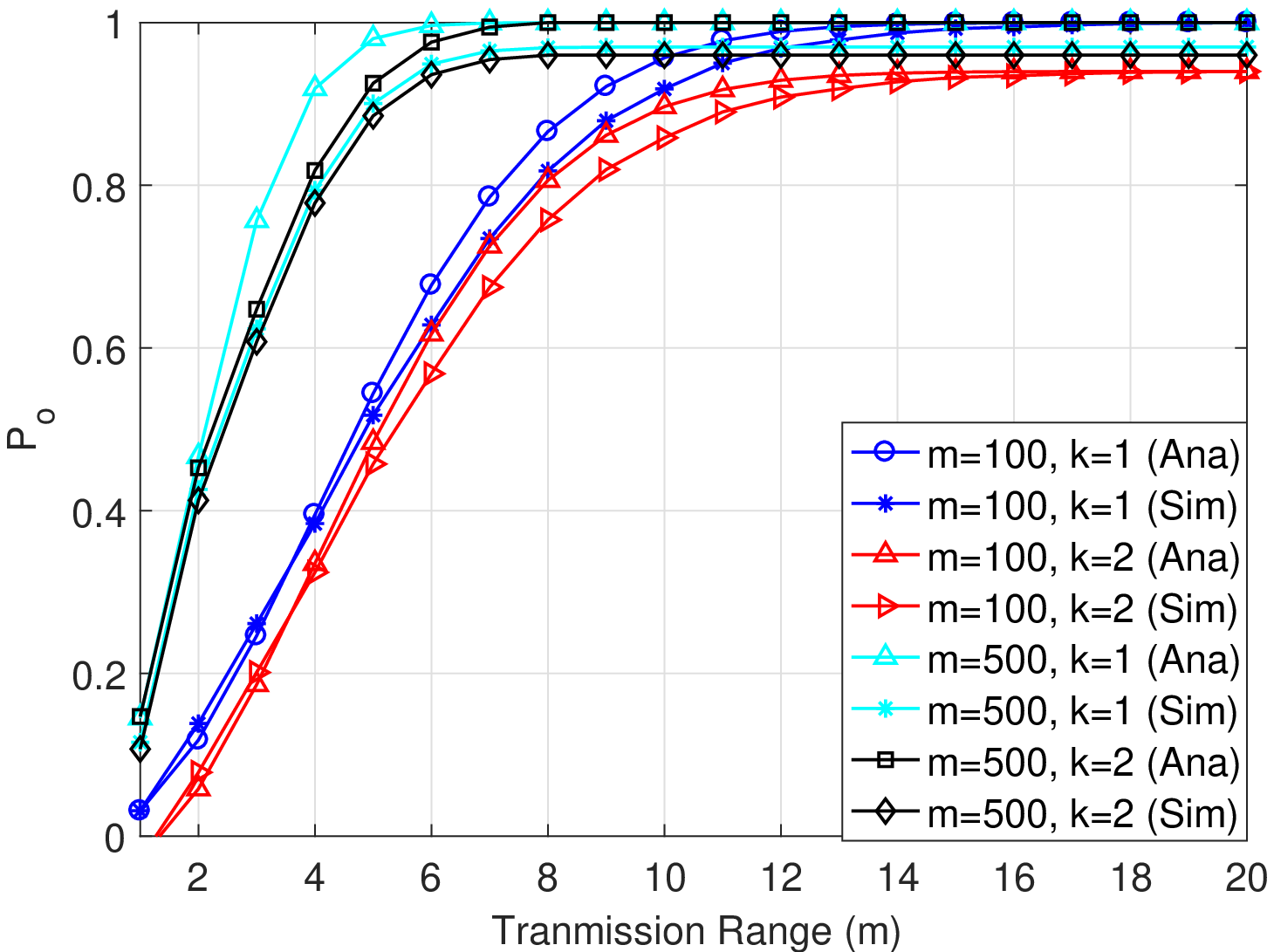}  
\caption{$p_o$ vs. transmission range for $\phi = 2\pi$.\label{fig:7}}  
\endminipage
\vspace{4pt}
\hrule
\end{figure*}

Since it is not always possible to have a link between the two nodes, $?$ terms in \eqref{eq: distance} denote the missing elements of the matrix $\mathbf{\hat{D}}$.  Even though recovery of observed distances for these missing terms may not be feasible, they can be replaced with any reasonable real positive values. It is well known that if $\mathit{O}(M^{1.2}\kappa\log(M))$ number of entries are available in  $\mathbf{\hat{D}}$ with rank $\kappa$, then the missing elements can be recovered \cite{Candes2012}. In order to approximate the low-rank matrix from the available entries in $\mathbf{\hat{D}}$, we consider the following optimization problem
\begin{subequations}
\begin{alignat}{2}
&\!\min_{\mathbf{\tilde{D}}\in \mathcal{R}^{M\times M}}        
&\qquad&  \text{rank}(\mathbf{\tilde{D}}) \label{eq: rank1}\\
& \hspace{15pt}\text{s.t.} &      & \mathsf{O}_{\Sigma}(\mathbf{\tilde{D}}) = \mathsf{O}_{\Sigma}(\mathbf{\hat{D}}),\label{eq:constraint1}
\end{alignat}
\end{subequations}
where $\mathsf{O}_\Sigma$ is the directed adjacency operator and is defined for any matrix $\mathbf{{A}}$  as
\begin{equation}
\mathsf{O}_\Sigma(\mathbf{{A}})_{ij}=
\begin{cases}
a_{ij}~~~\text{if}~ (i,j)\in\Sigma\\
0 ~~~~~\text{otherwise},
\end{cases}
\end{equation} 
and $\Sigma = \{(i,j): \mathbf{c}_j \in S_i ~\text{or}~ \mathbf{c}_i \in S_j\}$ are the set of indices. Note that the rank function in \eqref{eq: rank1} is nonlinear and non-convex, therefore it is numerically not possible to solve it. An alternative method is proposed in \cite{WEN2012} to solve the problem in  \eqref{eq: rank1} by using least square minimization which is formulated as follows
\begin{subequations}
\begin{alignat}{2}
&\!\min_{\mathbf{\tilde{D}}\in \mathcal{R}^{M\times M}}        
&\qquad&  \frac{1}{2}\parallel\mathsf{O}_{\Sigma}(\mathbf{\tilde{D}})- \mathsf{O}_{\Sigma}(\mathbf{\hat{D}})\parallel_F^2 \label{eq: rank2}\\
& \hspace{15pt}\text{s.t.} &      & \text{rank}(\mathbf{\tilde{D}})\leq \delta, \label{eq:constraint2}
\end{alignat}
\end{subequations}
where $\parallel.\parallel_F^2$ is the Forbenius norm and $\delta$ represents the upper bound for the rank. The expression in \eqref{eq: rank2} is a more effective approach to solve the optimization problem by relaxing the equality constraint in \eqref{eq:constraint1}. It is well known that when the number of nodes $M$ are distributed in $z$ dimensional space (i.e., $z=2$ for two-dimensional and $z=3$ for three-dimensional space) and $M \geq z$, then $\text{rank}(\mathbf{\hat{D}})\leq z+2$. Therefore, \eqref{eq: rank2} can be re-written as
\begin{subequations}
\begin{alignat}{2}
&\!\min_{\mathbf{\tilde{D}}\in \mathcal{R}^{M\times M}}        
&\qquad&  \frac{1}{2}\parallel\mathsf{O}_{\Sigma}(\mathbf{\tilde{D}})- \mathsf{O}_{\Sigma}(\mathbf{\hat{D}})\parallel_F^2 \label{eq: rank3}\\
& \hspace{15pt}\text{s.t.} &      & \text{rank}(\mathbf{\tilde{D}})\leq z+2. \label{eq:constraint3}
\end{alignat}
\end{subequations}
Since $\mathbf{\tilde{D}}$ is sparse, positive semi-definite, and the rank condition in \eqref{eq:constraint3} is always true. The problem in \eqref{eq:constraint3} can be solved by using conjugate gradient method \cite{Zhi2016}. The equation for updating the elements in $\mathbf{\tilde{D}}$ is given by
\begin{equation}\label{eq: update}
\mathbf{\tilde{D}}_{i+1}=\alpha_z(\mathbf{\tilde{D}}_{i}+\lambda \mathbf{T}_i)
\end{equation}
where $\alpha_z$ is the retraction operation which tells us about the direction of the tangent space while staying on the manifold, $\lambda$ is the step size, and $\mathbf{T}_i$ is direction of the function. $\mathbf{T}_i$ is computed as
\begin{equation}
\mathbf{T}_i = -\text{grad}f(\mathbf{\tilde{D}}_i)+\beta_i P(\mathbf{T}_{i-1}),
\end{equation}
where $\text{grad}f(\mathbf{\tilde{D}}_i)$ is the Riemannian conjugate gradient, $\beta_i$ is the tangent vector, and $P(\mathbf{T}_{i-1})$ is the orthogonal projection on the tangent space. Once the missing elements are recovered in $\mathbf{\tilde{D}}$ using the update rule in \eqref{eq: update}, the surface station is able to find out the relative position estimation of each node.

\subsection{Relative Position Estimation}
As the missing distances are estimated in the previous section, now the surface station uses data analysis methods to estimate the actual location of nodes. Some of the most famous data analysis methods are multidimensional scaling \cite{nasir2015, Rajawat2017}, Isomap \cite{Kashniyal2017}, and principal component analysis \cite{Nandakishore1997}. All of these methods are also called dimensionality reduction methods which tries to embed a higher dimensional data into a lower dimensional space. Following are the common steps in dimensionality reduction methods to estimate the lower dimensional position estimation from the higher dimensional observed euclidean distances $\mathbf{\tilde{D}}$.
\begin{itemize}
\item Compute the squared distance observation matrix $\boldsymbol{S} = \mathbf{\tilde{D}}^2$.

\item Double center $\boldsymbol{S}$ (i.e., $\boldsymbol{C} = −\frac{1}{2}\boldsymbol{JSJ}$) by using the centering operator $\boldsymbol{J = I}-M^{-1}\boldsymbol{11'}$, where $M$ are the total number of nodes, $\boldsymbol{I}$ is identity matrix of size $M \times M$, and $\boldsymbol{1}$ is an $M \times 1$ vector of ones. In double centering method the column and row means of a matrix are subtracted from each element of the matrix and its grand mean is added to each element.
\item Decompose matrix $\boldsymbol{C}$ and extract the $M$ eigenvectors $
\{\boldsymbol{e}_1, . . .,\boldsymbol{e}_M\}$ and the corresponding eigenvalues $\{\lambda_1,. . .,\lambda_M\}$. To get the two dimensional representation of nodes, only the largest two vectors $\boldsymbol{E}=\{\boldsymbol{e}_1, \boldsymbol{e}_2\}$ and the corresponding eigenvalues $\boldsymbol{\Lambda} = \{\lambda_1,\lambda_2\}$ are considered where the size of $\boldsymbol{E}$ is $M \times 2$ and $\boldsymbol{\Lambda}$ is $2 \times 2$ respectively.

\item Finally the position estimation of all the nodes with respect to their neighbors in a two dimensional space is given as
\begin{equation}\label{relative}
\boldsymbol{\hat{P}} = {\frac{1}{2}}\sqrt{\boldsymbol{\Lambda}}\boldsymbol{E}.
\end{equation}
\end{itemize}
\subsection{Final Position Estimation}
The relative position estimates are usually transformed into the global position estimation with the help of anchors. The transformation factors such as rotation, translation, and scaling are computed for the anchors based on their relative position estimates and actual locations. The rotation $\boldsymbol{\rho}$, translation $\boldsymbol{\tau}$, and scaling $s$ factors are computed by using orthogonal Procrustes analysis or Helmert transformation \cite{Alee2017}. 
The final position estimation primarily depends on the transformation factors of Procrustes analysis, i.e., 
\begin{equation}\label{finalpos}
\boldsymbol{\tilde{P}} = \boldsymbol{\rho} s(\boldsymbol{\hat{P}})+\boldsymbol{\tau}.
\end{equation}
where  $\boldsymbol{\rho}$ is rotation, $\boldsymbol{\tau}$ is translation, and $s$ is the scaling factor.  The rotation, translation, and scaling factors depend on the number of anchors and position of anchors. Consider that the actual position of anchors is $\boldsymbol{P}_a = \{\boldsymbol{p}_1,\boldsymbol{p}_2,...\boldsymbol{p}_k\}$ and their relative positions are $\boldsymbol{P}_r=\{\boldsymbol{\hat{p}}_1,\boldsymbol{\hat{p}}_2,...\boldsymbol{\hat{p}}_k\}$, then the objective function for Procrustes analysis is defined as
\begin{eqnarray}\label{eq: pro}
f(\boldsymbol{\rho},\boldsymbol{\tau},s)&=&\sum_{i=1}^k(\boldsymbol{\hat{p}}_i-s \boldsymbol{\rho}^T\boldsymbol{p}_i-\boldsymbol{\tau})^T\nonumber\\
& & \times (\boldsymbol{\hat{p}}_i-s \boldsymbol{\rho}^T\boldsymbol{p}_i-\boldsymbol{\tau})
\end{eqnarray}
The optimal values of $\boldsymbol{\rho}$,  $\boldsymbol{\tau}$, and  $s$ is obtained by minimization of \eqref{eq: pro}. Detailed derivations on the minimization of \eqref{eq: pro} are given in Appendix C.  Note that increasing the number of anchor nodes reduces the transformation error in \eqref{eq: pro} which reduces the localization error. However, for a certain number of anchor nodes the transformation error is almost zero after which the increase in number of nodes do not improve the localization error. The overall algorithm for the proposed localization system is summarized in Algorithm \ref{alg:main}.
\begin{algorithm}
\caption{Proposed localization method.}
{\bf Input}:  {Pairwise single hop noisy distances $\hat{d}_{ij}$ and set of anchors} \\
{\bf Outpu}t: {Position estimation of all nodes, i.e., $\boldsymbol{\tilde{P}}$ } \\
1: Compute matrix $\boldsymbol{\hat{D}}$ by using \eqref{eq: distance}  \\
2: Estimate the missing pairwise distance by using  \eqref{eq: update} \\
3: Estimate the relative position of each node by using  \eqref{relative}\\
4: Transformation to the final global position by using  \eqref{finalpos}\\
5: {\bf return}: Position estimations $\boldsymbol{\tilde{P}}$
\label{alg:main}
\end{algorithm}
\section{Numerical Results and Discussions}
\label{sec:results}
In this section, the connectivity analysis is numerically investigated with respect to different network parameters and results are compared to the theoretical derivations. Also, the performance of the proposed localization system is evaluated in terms of connectivity, robustness, and number of anchor nodes.

\subsection{Evaluation of UOWSNs Connectivity}
Simulations are conducted over a multi-hop UOWSN, deployed in an area of $100 m \times 100 m$ where $M$ optical sensor nodes are randomly distributed with random orientation between $0$ and $2 \pi$. The probability of having a connected network (i.e., $p_o=1$ when $k=1$ and $k=2$) is analytically and numerically evaluated in Fig. \ref{fig:4} - Fig. \ref{fig:7} which are obtained by averaging over 1000 network scenarios. The adjacency matrix $\boldsymbol{\Sigma}$ of each network scenario is obtained based on the selected parameters to observe the neighborhood relationships. 
\begin{figure}[ht]
\begin{center}
\includegraphics[width=0.95\columnwidth]{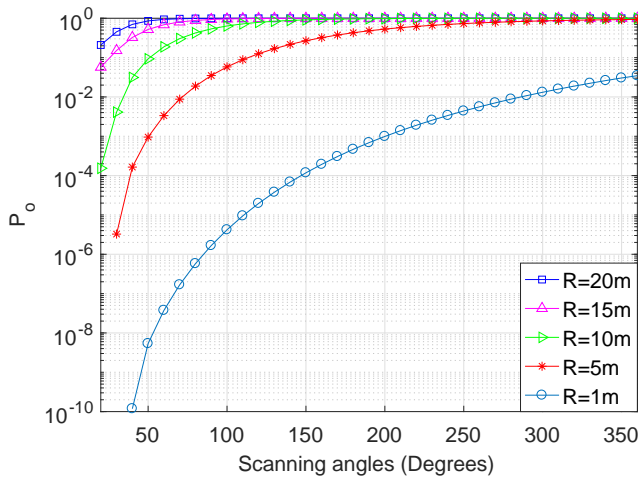}  
\caption{$p_o$ vs. Scanning angles for $M=10$.\label{fig:9}} 
\end{center}
\vspace{-1.5 em} 
\end{figure}
Through Fig. \ref{fig:4} - Fig. \ref{fig:7}, we set the beam scanning angles of the nodes with different widths of $\phi = \frac{2\pi}{9}, \frac{\pi}{2}, \frac{3\pi}{4}$, and $2\pi$ to see the impact of scanning angles on probability of a connected network.  The transmission range $R$ varies from 1 to 20 meters and the number of nodes $M$ are kept 100 and 500 respectively. As it is expected, the probability of connectivity increases as the number of nodes, range, and width of the scanning angles increases. In Fig. \ref{fig:4} - Fig. \ref{fig:7}, the analytically derived $p_o$ is larger than the simulated $p_o$ in all simulation settings because of the border effects where the nodes at the border become more isolated. Additionally, the border effects have more impact when the scanning angle is small (see Fig.~\ref{fig:4} when $\phi =\frac{2\pi}{9}$) resulting in more isolated nodes at the border. However, increasing the transmission range, the number of nodes and beam scanning angles reduces the border effects which scale down the difference between the analytically derived solution and the simulated results as shown in Fig. \ref{fig:7}. One can deduce that number of nodes, range, and width of the scanning angles are the most critical design parameters for a connected UOWSN.

In the second simulation scenario, a sparse network is considered with $M=10$ in a square region of $100 m \times 100 m$ where the effect of transmission range and the beam scanning angle is investigated. It is clear from Fig.~\ref{fig:9} that the probability of no obscured node  $p_o = 0.99$ cannot be achieved even in omnidirectional transmission case ($\phi=2\pi$) with a transmission range of 1 m. But when the transmission range is increased to 10 m and 20 m the connectivity is achieved at $\phi=\frac{\pi}{2}$ and $\phi=\frac{2\pi}{9}$, respectively. Thus, careful consideration of $M$, $\phi$ and $R$ are important parameters for deployment of a connected multi-hop UOWSN. We conclude this section by summarizing the following insights:
\begin{itemize}
\item 
The linearly increase of transmission range $R$ have a more noticeable effect on $p_o$ than linear increase in $\phi$.
\item 
Analogous to the random geometric graphs, there is a particular value of transmission range $R$ above which there is almost no obscured node in the network. Also, the phase transition region of $p_o$ versus transmission range increases as $\phi$ becomes smaller.
\item In case of $k$-connectivity, as $k$ grows, it will require significantly large values of $R$ to keep the network connected.
\item 
This paper demonstrates a practical deployment solution for a connected UOWSNs, where connectivity is one of the major hurdles to deploy UOWSNs. The potential to guarantee connectivity in UOWSNs further supports the deployment of emerging underwater directional optical wireless systems. 
\end{itemize} 
\begin{figure}
\begin{center}
\includegraphics[width=0.95 \columnwidth]{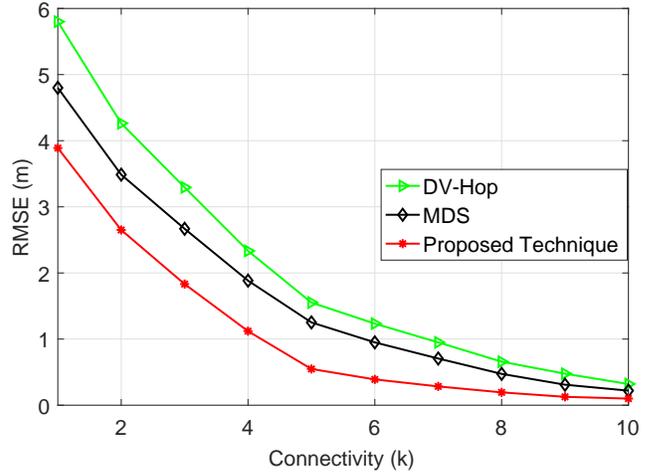}  
\caption{RMSE vs. Connectivity for $M=100$.\label{fig:10}} 
\end{center}
\vspace{-1.5 em} 
\end{figure}
\subsection{Evaluation of Localization Performance}
In this section, we evaluate the performance of the proposed localization systems in terms of connectivity, robustness, and number of anchor nodes. Also, we compare the results with other well-known network localization schemes such as multidimensional scaling (MDS) \cite{Shang} and distance vector routing (DV) \cite{Niculescu.D}.
\subsubsection{Impact of Connectivity}
To check the impact of connectivity on localization error, we have simulated a scenario with 100 optical sensor nodes and 5 anchor nodes randomly deployed in  $100 m \times 100 m$ square area. Fig.~\ref{fig:10} shows the impact of connectivity on root mean square error (RMSE) performance, where RMSE is defined as
\begin{equation}
\text{RMSE} = \frac{\sqrt{\sum_{i=1}^M(\boldsymbol{p}_i-\tilde{\boldsymbol{p}}_i)^2}}{M},
\end{equation}
where $\tilde{\mathbf{p}}_i$ is the estimated location of node $i$. In Fig.~\ref{fig:10} we have compared the RMSE performance of the proposed technique with other well-known network localization schemes such as multidimensional scaling (MDS) \cite{Shang} and distance vector routing (DV) \cite{Niculescu.D}. It is clear from Fig.~\ref{fig:10} that the proposed technique outperforms the literature because it approximates the missing distances more accurately in the observation distance matrix. Also, the RMSE performance improves with an increase in the connectivity of the network.
\subsubsection{Robustness}
\begin{figure}
\begin{center}
\includegraphics[width=0.95 \columnwidth]{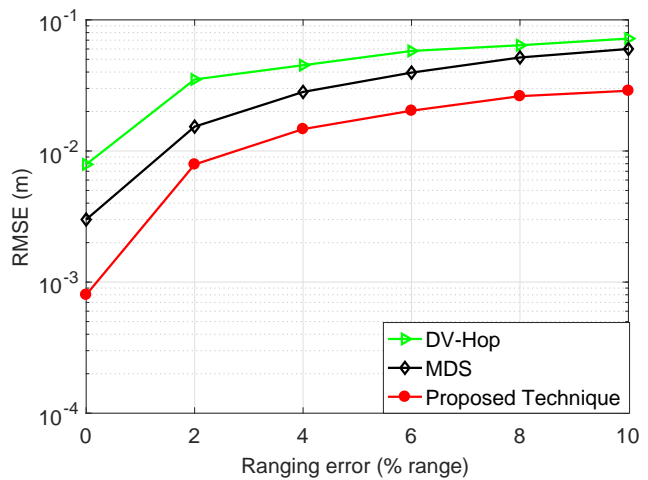}  
\caption{RMSE vs. Ranging error.\label{fig:11}} 
\end{center} 
\end{figure}

Undoubtedly, the ranging error and the estimation of missing pairwise distances have a negative effect on the accuracy of every localization system. Here, we examine the performance of the proposed technique in the presence of ranging error. To examine the impact of ranging error we considered 100 optical sensor nodes and 5 anchor nodes randomly deployed in  $100 m \times 100 m$ square area with the transmission range of 40 m and beam scanning angle of $\frac{3 \pi}{4}$. Assuming that the ranging errors are Gaussian distributed with zero mean and variance $\sigma^2$, where the values of $\sigma^2$ are set to 2-10 \% of the range. Note that the results are averaged our 100 different network setups. Fig.~ \ref{fig:11} shows that the proposed localization system is more robust to the ranging error as compared to MDS and DV hop.  This is mainly because of the better approximation of missing pairwise distances.
\subsubsection{Number of Anchors}
\begin{figure}
\begin{center}
\includegraphics[width=0.95\columnwidth]{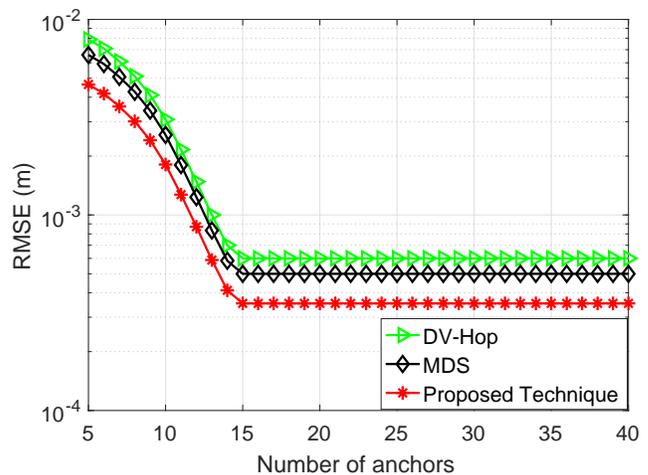}  
\caption{RMSE vs. Number of anchors.\label{fig:12}} 
\end{center} 
\vspace{-1.5 em}
\end{figure}

It is a matter of fact that increasing the number of anchors for localization systems improves the RMSPE up to a certain extent. To evaluate the performance of the proposed method in connection to the number of anchors,   we have simulated a scenario with 100 optical sensor nodes and a variable number of anchor nodes randomly deployed in  $100 m \times 100 m$ square area. Fig.~\ref{fig:12} shows that increasing the number of nodes up to 15 improves RMSPE. When the number of anchor nodes is 15 the network becomes saturated and a further increase in the number of anchor nodes do not improve RMSPE.
\section{Conclusions and Future Works}
In this paper, a stochastic network connectivity is analyzed for multi-hop UOWSNs from a graph theoretical approach. In this regard, UOWSNs are modeled as uni-directional random sector graphs where the coverage region of each node is angular-sector shaped due to the propagation characteristics of the light beam. Based on descendant and antecedent neighbor definitions, a closed-form expression of the probability of $k$-connectivity is derived for UOWSNs as a function of beam scanning angle, the number of nodes, and transmission range. Throughout extensive simulations, we show that our analytical findings comply with simulations and  UOWSNs parameters play a key role to achieve a desirable network connectivity. Furthermore, the network localization is formulated as an unconstrained optimization problem and solved using the conjugate gradient technique. Numerical results of the proposed localization method are compared with the literature where the localization performance of the proposed method outperforms the literature in terms of network connectivity, ranging error, and the number of anchors. Large-scale UOWSNs preclude the use the centralized localization methods where reporting the pairwise range measurements to a centralized node from each sensor node and then sending back the estimated location to the node overwhelm the capacity of UOWSNs and waste energy. Therefore, it is required to develop a distributed version of the proposed localization method to balance the communication and computational load over all the nodes in the network. Additionally, the proposed method estimates the position of underwater sensors in two-dimensional space. It would be interesting to extend the proposed localization method to its three-dimensional counterpart to better realize the underwater environment.
\label{sec:conc}
\section{Acknowledgement}
The authors would like to thank the anonymous reviewers for
their fruitful comments.
\appendices
\section{Derivation of $p^i_{a|d}$}
Simplifying \eqref{eq: jointprob2} we obtain 
\begin{eqnarray} \label{eq: jointprob4}
\nonumber p^i_{a|d} &=& 1-\text{Pr}[A_i = 0 | D_i \geq 1, \Upsilon = 0] \text{ Pr}[\Upsilon = 0]\\
& = & 1- \sum_{b=1}^{M-1}\text{Pr}[A_i = 0 | D_i = b, \Upsilon = 0] \times \nonumber\\
& & \text{Pr}[\Upsilon = 0| D_i = b] \text{ Pr}[D_i = b |D_i \geq 1],
\end{eqnarray}
where index $b$ represents the bi-directional links and summation terms are given by
\begin{equation}\label{eq: jointprob42}
\text{Pr}[A_i = 0 | D_i = b, \Upsilon = 0] = \left(1-\frac{\phi R^2}{2}\right)^{M-b-1},
\end{equation}
\begin{equation}\label{eq: jointprob41}
\text{Pr}[\Upsilon = 0| D_i = b] = \left(1-\frac{\phi}{2\pi}\right)^b, \;1 \leq b \leq M, \text{ and}
\end{equation}
\begin{equation}\label{eq: jointprob43}
\text{Pr}[D_i = b |D_i \geq 1] = \frac{\exp\{Q\}Q^b}{b!\left(1-\exp\{Q\}\right)}.
\end{equation}
Substituting \eqref{eq: jointprob42}-\eqref{eq: jointprob43} into \eqref{eq: jointprob4}, we obtain the conditional backward connectivity $p^i_{a|d} $ as follows
\begin{eqnarray}\label{eq: jointprobfinal0}
p^i_{a|d} & = & 1- \sum_{b=1}^{M-1} \left(1-\frac{\phi R^2}{2} \right)^{M-b-1} \left(1-\frac{\phi}{2\pi} \right)^b \nonumber \\
& &  \frac{\exp\{Q\}Q^b}{b!\left(1-\exp\{Q\}\right)}, \forall i.
\end{eqnarray}
When there exists only a single bi-directional link, i.e., $b=1$, \eqref{eq: jointprobfinal0} can be further simplified as
\begin{eqnarray}\label{eq: jointprobfinal1}
p^i_{a|d}  &=& 1-\frac{\exp\{Q\}}{1-\exp\{Q\}}\left(1-\frac{\phi R^2}{2}\right)^{M-1} \times \nonumber \\ & &\left(\exp \left \{ \frac{-Q (2\pi - \phi)}{\pi(2-\phi R^2)} \right \} -1 \right), \forall i.
\end{eqnarray}
\section{Derivation of $p^i_{a_{2}|d_{2}}$}
To derive the expression for $p^i_{a_{2}|d_{2}}$, we consider two events, first when node $n_i$ has one bi-directional link ($\Upsilon = 1$) and second when node $n_i$ has two bi-directional links ($\Upsilon = 2$):
\begin{eqnarray}\label{eq: kconnect}
p^i_{a_{2}|d_{2}} & = & \text{Pr}[A_i \geq 2 | D_i \geq 2 ] =  1-\text{Pr}[A_i < 2 | D_i \geq 2] \nonumber \\ & = & 1-\{\text{Pr}[A_i = 0 | D_i \geq 2]+\text{Pr}[A_i = 1 | D_i \geq 2]\} \nonumber \\ & = & 1-\left\{\text{Pr}[A_i = 0 | D_i \geq 2, \Upsilon = 0]\text{Pr}[\Upsilon = 0] \right. \nonumber \\
& & \left. + \text{Pr}[A_i = 1 | D_i \geq 2, \Upsilon = 0] \text{Pr}[\Upsilon = 0]\right\} \nonumber \\
& & -\left\{\text{Pr}[A_i = 0 | D_i \geq 2, \Upsilon = 1]\text{Pr}[\Upsilon = 1] \right. \nonumber \\
& & \left. + \text{Pr}[A_i = 1 | D_i \geq 2, \Upsilon = 1] \text{Pr}[\Upsilon = 1]\right\} \nonumber \\
& & - \text{Pr}[A_i = 1 | D_i \geq 2, \Upsilon = 2] \text{Pr}[\Upsilon = 2].
\end{eqnarray}
where the contradicting terms $\text{Pr}[A_i = 0 | D_i \geq 2, \Upsilon = 1]$ and $\text{Pr}[A_i = 1 | D_i \geq 2, \Upsilon = 2]$ are equal to zero. Thus, eliminating these zero terms in \eqref{eq: kconnect}  and expanding the other terms for $b = 2,3,...M-1$, we obtain $p^i_{a_{2}|d_{2}} $ as
\begin{eqnarray}\label{eq: kconnect2}
p^i_{a_{2}|d_{2}} & = & 1 - \text{Pr}[A_i = 0 | D_i \geq 2, \Upsilon = 0]\text{Pr}[\Upsilon = 0] \nonumber \\
& & - \text{Pr}[A_i = 1 | D_i \geq 2, \Upsilon = 0] \text{Pr}[\Upsilon = 0] \nonumber \\
& & - \text{Pr}[A_i = 1 | D_i \geq 2, \Upsilon = 1] \text{Pr}[\Upsilon = 1].
\end{eqnarray}
\begin{eqnarray}\label{eq: kconnect3}
p^i_{a_{2}|d_{2}} & = & 1 - \sum_{b=2}^{M-1}\left\{\text{Pr}[A_i = 0 | D_i =  b, \Upsilon = 0]\right. \nonumber \\ & & \left. \text{Pr}[\Upsilon = 0 | D_i =  b].\text{Pr}[D_i = b | D_i \geq  2]\right\} \nonumber \\
& & -\sum_{b=2}^{M-1}\left\{\text{Pr}[A_i = 1 | D_i =  b, \Upsilon = 0]\right. \nonumber \\ & & \left. \text{Pr}[\Upsilon = 0 | D_i =  b].\text{Pr}[D_i = b | D_i \geq  2]\right\} \nonumber \\
& & -\sum_{b=2}^{M-1}\left\{\text{Pr}[A_i = 1 | D_i =  b, \Upsilon = 1]\right. \nonumber \\ & & \left. \text{Pr}[\Upsilon = 1 | D_i =  b].\text{Pr}[D_i = b | D_i \geq  2]\right\}. 
\end{eqnarray}
To simplify, the three summation terms in \eqref{eq: kconnect3} are represented by $\mathbb{S}_1$, $\mathbb{S}_2$ and $\mathbb{S}_3$ respectively. The closed form solution for the probabilities in \eqref{eq: kconnect3} are given as
\begin{equation}\label{eq: k1}
\text{Pr}[A_i = 0 | D_i =  b, \Upsilon = 0] = \left(1-\frac{\phi R^2}{2\pi}\right)^{M-b-1},
\end{equation}
\begin{equation}\label{eq: k2}
\text{Pr}[\Upsilon = 0 | D_i =  b] = \left(1-\frac{\phi}{2\pi}\right)^{b},
\end{equation}
\begin{equation}\label{eq: k3}
\text{Pr}[D_i = b | D_i \geq  2] = \frac{-Q \exp\{Q\}}{b ! \left(1-\exp\{Q\}(1-Q)\right)},
\end{equation}
\begin{eqnarray}\label{eq: k4}
\text{Pr}[A_i & = & 1 | D_i =  b, \Upsilon = 0] = (M-b-1)\left(\frac{\phi R^2}{2}\right) \nonumber \\
& & \left(1-\frac{\phi R^2}{2}\right)^{M-b-2},
\end{eqnarray}
and
\begin{equation}\label{eq: k5}
\text{Pr}[A_i = 1 | D_i =  b, \Upsilon = 1] = \left(1-\frac{\phi R^2}{2\pi}\right)^{M-b-1}.
\end{equation}
Substituting \eqref{eq: k1}-\eqref{eq: k5} in \eqref{eq: kconnect3}, we obtain the three summation terms  $\mathbb{S}_1$, $\mathbb{S}_2$ and $\mathbb{S}_3$ as follows
\begin{eqnarray}\label{eq: s1}
\mathbb{S}_1 & = & \frac{\exp\{Q\}}{1-(1-Q)\exp\{Q\}}\left(1-\frac{\phi R^2}{2}\right)^{M-1} \nonumber \\
& & \left(\exp^{\frac{-Q(2\pi-\phi)}{\pi(2-\phi R^2)}}+\frac{Q(2\pi-\phi)}{\pi(2-\phi R^2)} - 1\right),
\end{eqnarray}
\begin{eqnarray}\label{eq: s2}
\mathbb{S}_2 & = & \frac{\exp\{Q\}}{1-(1-Q)\exp\{Q\}}\left(\frac{\phi R^2}{2}\right)\left(1-\frac{\phi R^2}{2}\right)^{M-2} \nonumber \\
& & \left[(M-1)\left(\exp^{\frac{-Q(2\pi-\phi)}{\pi(2-\phi R^2)}}+\frac{Q(2\pi-\phi)}{\pi(2-\phi R^2)}\right) \right.\nonumber \\
& & \left. - \frac{Q(2\pi-\phi)}{\pi(2-\phi R^2)}\left(\exp^{\frac{-Q(2\pi-\phi)}{\pi(2-\phi R^2)}}-1\right)-1\right],
\end{eqnarray}
and
\begin{eqnarray}\label{eq: s3}
\mathbb{S}_3 & = & \frac{\exp\{Q\}}{1-(1-Q)\exp\{Q\}} \left(\frac{M\phi^2 R^2}{4\pi}\right)\left(1-\frac{\phi R^2}{2}\right)^{M-2} \nonumber \\ 
& & \left(\exp^{\frac{-Q(2\pi-\phi)}{\pi(2-\phi R^2)}}-1\right).
\end{eqnarray}
Substituting \eqref{eq: s1}, \eqref{eq: s2} and \eqref{eq: s3} in \eqref{eq: kconnect3}, we obtain
\begin{eqnarray}\label{eq: finalconnect}
p^i_{a_{2}|d_{2}} & = & 1 - \frac{\exp\{Q\}}{1-(1-Q)\exp\{Q\}}\left(1-\frac{\phi R^2}{2}\right)^{M-1} \nonumber \\
& & \left(\exp^{\frac{-Q(2\pi-\phi)}{\pi(2-\phi R^2)}}+\frac{Q(2\pi-\phi)}{\pi(2-\phi R^2)} - 1\right) \nonumber \\
& & - \frac{\exp\{Q\}}{1-(1-Q)\exp\{Q\}}\left(\frac{\phi R^2}{2}\right)\left(1-\frac{\phi R^2}{2}\right)^{M-2} \nonumber \\
& & \left[(M-1)\left(\exp^{\frac{-Q(2\pi-\phi)}{\pi(2-\phi R^2)}}+\frac{Q(2\pi-\phi)}{\pi(2-\phi R^2)}\right) \right.\nonumber \\
& & \left. - \frac{Q(2\pi-\phi)}{\pi(2-\phi R^2)}\left(\exp^{\frac{-Q(2\pi-\phi)}{\pi(2-\phi R^2)}}-1\right)-1\right] \nonumber \\
& & -\frac{\exp\{Q\}}{1-(1-Q)\exp\{Q\}} \left(\frac{M\phi^2 R^2}{4\pi}\right) \nonumber \\ 
& & \left(1-\frac{\phi R^2}{2}\right)^{M-2} \left(\exp^{\frac{-Q(2\pi-\phi)}{\pi(2-\phi R^2)}}-1\right).
\end{eqnarray}
\bibliographystyle{../bib/IEEEtran}
\bibliography{../bib/IEEEabrv,../bib/nasir_ref}
\vspace{-2.3 em}
\begin{IEEEbiography}[{\includegraphics[width=1in,height=1.25in]{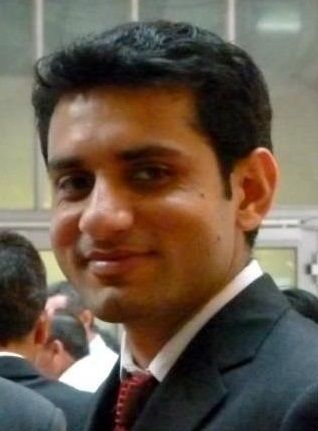}}]{Nasir Saeed}(S'14-M'16) received his Bachelors of Telecommunication degree from University of Engineering and Technology, Peshawar, Pakistan, in 2009 and received Masters degree in satellite navigation from Polito di Torino, Italy, in 2012. He received his Ph.D. degree in electronics and communication engineering from Hanyang University, Seoul, South Korea in 2015. He was an assistant professor at the Department of Electrical Engineering, Gandhara Institute of Science and IT, Peshawar, Pakistan from August 2015 to September 2016. Dr. Saeed worked as an assistant professor at IQRA National University, Peshawar, Pakistan from October 2017 to July 2017. He is currently a postdoctoral research fellow at Communication Theory Lab, King Abdullah University of Science and Technology (KAUST).   His current areas of interest include cognitive radio networks, underwater optical wireless communications, dimensionality reduction, and localization.
\end{IEEEbiography}

\begin{IEEEbiography}[{\includegraphics[width=1in,height=1.25in]{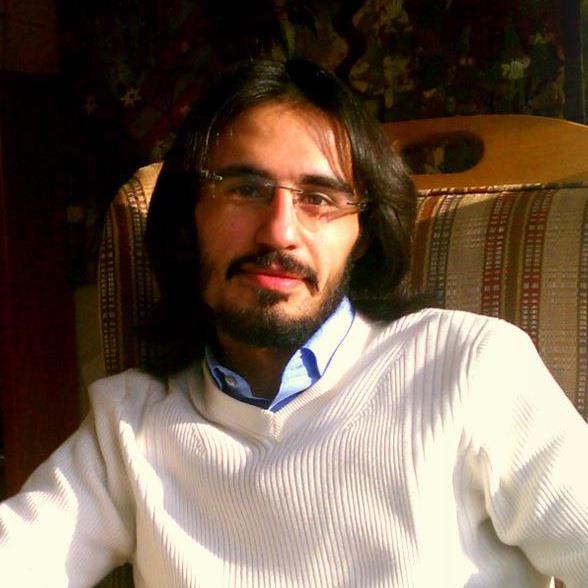}}]{Abdulkadir Celik}
(S'14-M'16) received the B.S. degree in electrical-electronics engineering from Selcuk University in 2009, the M.S. degree in electrical engineering in 2013, the M.S. degree in computer engineering in 2015, and the Ph.D. degree in co-majors of electrical engineering and computer engineering in 2016, all from Iowa State University, Ames, IA. He is currently a postdoctoral research fellow at Communication Theory Laboratory of King Abdullah University of Science and Technology (KAUST). His current research interests include but not limited to 5G networks and beyond, wireless data centers, UAV assisted cellular and IoT networks, and underwater optical wireless communications, networking, and localization. 
\end{IEEEbiography}
\vspace{-3.5 em}
\begin{IEEEbiography}[{\includegraphics[width=1in,height=1.25in]{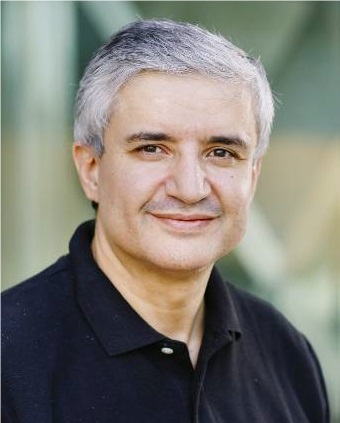}}]{Mohamed-Slim Alouini}
(S'94-M'98-SM'03-F'09)  was born in Tunis, Tunisia. He received the Ph.D. degree in Electrical Engineering
from the California Institute of Technology (Caltech), Pasadena,
CA, USA, in 1998. He served as a faculty member in the University of Minnesota,
Minneapolis, MN, USA, then in the Texas A\&M University at Qatar,
Education City, Doha, Qatar before joining King Abdullah University of
Science and Technology (KAUST), Thuwal, Makkah Province, Saudi
Arabia as a Professor of Electrical Engineering in 2009. His current
research interests include the modeling, design, and
performance analysis of wireless communication systems.
\end{IEEEbiography}
\begin{IEEEbiography}[{\includegraphics[width=1in,height=1.25in]{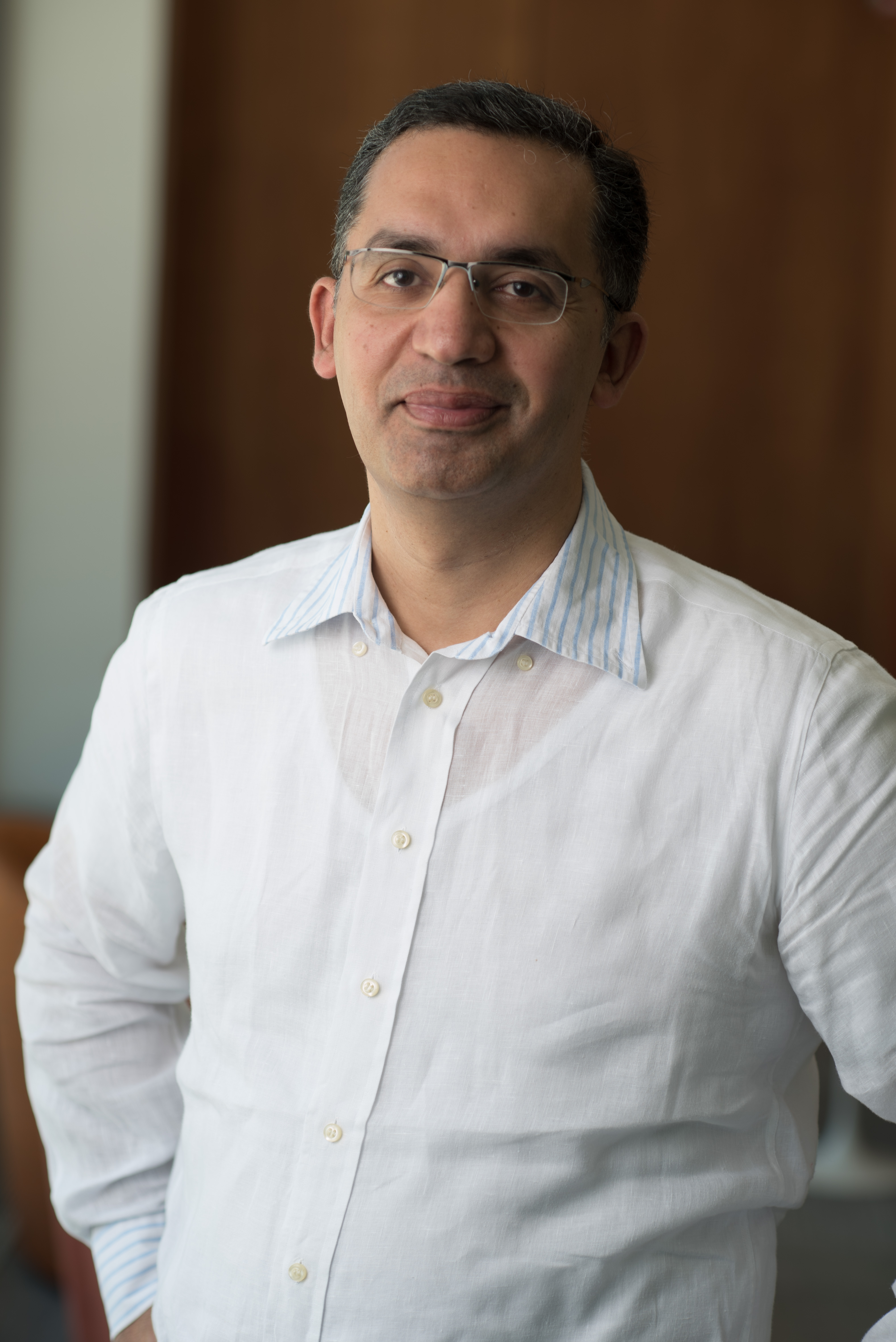}}]{Tareq Y. Al-Naffouri }
(M'10-SM'18) Tareq  Al-Naffouri  received  the  B.S.  degrees  in  mathematics  and  electrical  engineering  (with  first  honors)  from  King  Fahd  University  of  Petroleum  and  Minerals,  Dhahran,  Saudi  Arabia,  the  M.S.  degree  in  electrical  engineering  from  the  Georgia  Institute  of  Technology,  Atlanta,  in  1998,  and  the  Ph.D.  degree  in  electrical  engineering  from  Stanford  University,  Stanford,  CA,  in  2004.  

He  was  a  visiting  scholar  at  California  Institute  of  Technology,  Pasadena,  CA  in  2005  and  summer  2006.  He  was  a  Fulbright scholar  at  the  University  of  Southern  California  in  2008.  He  has  held  internship  positions  at  NEC  Research  Labs,  Tokyo,  Japan,  in  1998,  Adaptive  Systems  Lab,  University  of  California  at  Los  Angeles  in  1999,  National  Semiconductor,  Santa  Clara,  CA,  in  2001  and  2002,  and  Beceem  Communications  Santa  Clara,  CA,  in  2004.  He  is  currently  an  Associate Professor  at  the  Electrical  Engineering  Department,  King  Abdullah  University  of  Science  and  Technology  (KAUST).  His  research  interests  lie  in  the  areas  of  sparse, adaptive,  and  statistical  signal  processing  and  their  applications,  localization,  machine  learning,  and  network  information  theory.    He  has  over  240  publications  in  journal  and  conference  proceedings,  9  standard  contributions,  14  issued  patents,  and  8  pending. 

Dr.  Al-Naffouri  is  the  recipient  of  the  IEEE  Education  Society  Chapter  Achievement  Award  in  2008  and  Al-Marai  Award  for  innovative  research  in  communication  in  2009.  Dr.  Al-Naffouri  has  also  been  serving  as  an  Associate  Editor  of  Transactions  on  Signal  Processing  since  August  2013. 
\end{IEEEbiography}

\end{document}